\documentclass[a4paper,11pt]{article}
\usepackage{jheppub}
\usepackage{lineno}

\usepackage{bm,amssymb,amsmath,dsfont,mathrsfs,amstext,latexsym,physics,relsize}
\usepackage{graphicx}
\usepackage{mathtools}

\usepackage[dvipsnames]{xcolor}
\usepackage[numbers,sort&compress]{natbib}
\usepackage[colorlinks=true,citecolor=blue,urlcolor=blue,linkcolor=blue]{hyperref}
\usepackage[all]{hypcap}
\usepackage{comment}
\newcommand{\mycomment}[1]{}
\usepackage{soul}
\usepackage{tikz}

\let\Re\relax
\let\Im\relax
\DeclareMathOperator{\Re}{Re}
\DeclareMathOperator{\Im}{Im}

\newcommand{\figpanel}[2]{\hyperref[#1]{\ref{#1}#2}}

\newcommand{\orcidicon}[1]{\href{https://orcid.org/#1}{\includegraphics[height=1.6ex]{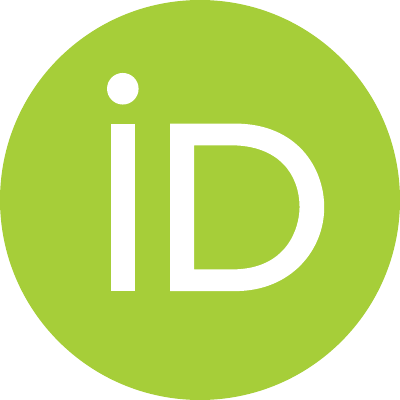}}}

\title{\boldmath Krylov Dynamics and Operator Growth in Time-Dependent Systems via Lie Algebras}

\author[1]{Andr\'as Grabarits\,\orcidicon{0000-0002-0633-7195}}
\author[2]{E.~Medina-Guerra\,\orcidicon{0000-0003-1040-1812}}
\author[1,3]{Adolfo del Campo\,\orcidicon{0000-0003-2219-2851}}

\affiliation[1]{Department of Physics and Materials Science, University of Luxembourg, L-1511 Luxembourg, Luxembourg}
\affiliation[2]{Department of Condensed Matter Physics, Weizmann Institute of Science, Rehovot 7610001, Israel}
\affiliation[3]{Donostia International Physics Center, E-20018 San Sebasti\'an, Spain}

\emailAdd{andras.grabarits@uni.lu}
\abstract{
We study quantum dynamics generated by time-dependent Hamiltonians in Krylov space, the minimal subspace in which the evolution takes place. We establish a direct link between dynamics in the time-dependent Krylov subspace and the underlying Lie-algebraic structure of the Hamiltonian. Exploiting the geometry of the root system, we develop a general framework in which the dynamics in the time-dependent Krylov subspace is generated by ladder operators of the associated Lie algebra. In particular, we identify the minimal conditions under which the exact time-dependent Krylov dynamics is naturally determined by the interaction-picture Hamiltonian and governed by an embedded $\mathfrak{sl}(2,\mathbb{C})$ subalgebra. We further show that an exact single-exponential representation of the time-evolution operator gives rise to a distinct time-independent Krylov dynamics in a unitarily related basis, from which the exact time-dependent Krylov dynamics can nevertheless be recovered. We also extend the framework to the oscillator algebra as the simplest extension of the nilpotent Heisenberg--Weyl algebra, and provide further examples, including the translated and dilated harmonic oscillator, systems governed by closed Virasoro subalgebras, a spin in a rotating magnetic field, and higher-dimensional generalizations for multi-level systems. In addition, we introduce a new quantum speed limit to the complexity growth rate generated by a time-dependent generator and show that, for evolutions governed by a Lie algebra, it retains the same functional form as in the time-independent case. Remarkably, saturation of this bound is strongly affected by temporal driving and persists only when the Hamiltonian commutes with itself at different times. These results establish a unified framework for characterizing operator growth and Krylov complexity in time-dependent quantum systems with underlying Lie-algebraic structures.
}

\begin{document}
\maketitle
\flushbottom


\section{Introduction}

Understanding the complexity of quantum dynamics remains a central challenge in modern physics. The exponential growth of Hilbert space makes the description of quantum evolution increasingly difficult as the number of degrees of freedom increases. Developing frameworks that capture essential features of quantum dynamics while reducing this complexity is therefore of fundamental importance. In this context, Krylov subspace methods have emerged as a versatile framework for analyzing the structure of quantum dynamics~\cite{Parker_2019,Nandy25,Baiguera2026,Rabinovici2025}. By representing the dynamics generated by a Hamiltonian in a dynamically constructed Krylov basis, these methods provide a geometric picture of quantum evolution~\cite{Caputa2021_Geometry_KC,Vijay2022_KrylovLadder} and have recently found applications in the study of operator growth, quantum chaos, delocalization in Krylov space,  quantum complexity and related measures ~\cite{DymarskySmolkin2021,DymarskyGorsky2020,RabinoviciSanchezGarridoShirSonner2022,Bhattacharjee2022,AvdoshkinDymarskySmolkin2024,HornedalEtAl2022,Craps2024,Nandy_2024,Nandy25,Takahashi2026,Choudhary2026}. They provide a fundamental description of quantum dynamics and have further found applications to quantum control \cite{Hornedal2023geometricoperator,Takahashi2024,Bhattacharjee2023,Hatomura_2024,Morawetz2025,Grabarits25_KrylovCD}, the study of critical phenomena \cite{Bento2024,Grabarits2025,Takahashi2025}, and quantum optimization algorithms \cite{Visuri2026}. 

Despite this progress, most existing formulations of the Krylov framework are restricted to time-independent generators. They are thus concerned with the preparation of an initial stationary state and its subsequent time-evolution, i.e., following a sudden quantum quench. In this setting, Krylov methods have also led to quantum speed limits for the growth of Krylov complexity~\cite{HornedalEtAl2022}, providing a direct bound in terms of the Liouvillian variance and the spread of the Krylov wave function. Evolutions saturating the Krylov complexity growth are described in terms of generalized coherent states.  By contrast, for explicitly time-dependent generators, the only available speed limit in Krylov space to date addresses the non-escape probability in the Krylov lattice ~\cite{takahashi2024_TDKC}, providing a complementary diagnostic tool in the spirit of Lieb-Robinson bounds. Such a bound focuses on the full probability distribution over the Krylov basis, and is generally more involved than the bound on the complexity growth rate.

Extending the ideas above to explicitly time-dependent systems remains considerably more challenging. In general, Hamiltonians at different times do not commute, and the resulting evolution operator involves time ordering, obscuring the simple geometric picture underlying the Krylov construction. 
Developing such a framework is crucial for uncovering how complexity builds up in driven quantum systems and for characterizing the spread of chaos and scrambling of information in nonequilibrium dynamics. 
Recent progress has extended Krylov methods to time-dependent generators. In the case of periodic systems and quantum circuits, such an extension is possible considering powers of an elementary unitary~\cite{Mitra2021,Yates2022,Nizami_2023,Nizami2024,Mitra2024,Suchsland2025}. For generic time-dependent systems, one approach is to use the exponential of the Floquet operator, which is essentially nonlocal in time \cite{takahashi2024_TDKC}. 
This approach allows the Krylov complexity to be tackled in time-dependent solvable systems. Its application is more challenging in numerical simulations of many-body systems, as the Floquet operator involves a time derivative.
An alternative approach is to 
express the time evolution unitary as the exponential of the Magnus operator, thus keeping a structural form analogous to that in the time-independent case \cite{Grabarits2025}. 

The presence of symmetries in physical systems is expected to facilitate the study of quantum complexity. Specifically, dynamical symmetries often allow for an exact description of quantum evolution and thus have wide-ranging applications ranging from adiabatic theory \cite{Lewis_Riesenfeld_1969,Mostafazadeh2001} to particle production in field theory \cite{Birrell1982} and quantum control \cite{TORRONTEGUI2013_STA}. 
In this work, we develop a Lie-algebraic construction for nonequilibrium time-evolution dynamics in Krylov space. In particular, for time-dependent Hamiltonians with an underlying simple Lie algebra, we identify the minimal conditions under which the dynamics in the interaction picture is generated directly by ladder operators of an embedded $\mathfrak{sl}(2,\mathbb{C})$ subalgebra, leading to a generalized coherent-state structure~\cite{Perelomov1972,Perelomov_1977,Perelomov1986,ban1993decomposition,Zhang1990,Vijay2022_KrylovLadder,Patramanis2022}. This construction naturally captures dynamics in a higher-dimensional Krylov lattice associated with the root structure of the underlying Lie algebra, and connects the corresponding factorized time-evolution operator to Wei-Norman factorizations~\cite{WeiNorman1963,Charzynski_2013,ALTAFINI2005}. This approach is further extended to the Heisenberg--Weyl algebra and its oscillator extension, which provide a natural nilpotent setting for physical applications. We further show that applying the Lanczos algorithm to the exact single-exponential generator of the time-evolution operator yields a distinct effective time-independent Krylov dynamics in a fictitious time. This auxiliary dynamics generally contains nonzero diagonal Lanczos coefficients and reproduces the exact time-dependent Krylov wavefunction when evaluated at unit fictitious time. 
Finally, we introduce a quantum speed limit to the Krylov complexity growth induced by a time-dependent generator and analyze it in the corresponding algebraic setting, where the bound retains the same Robertson form as in the time-independent case~\cite{HornedalEtAl2022}. Its saturation, however, is affected in an essential way by the driving. While the lowest-weight coherent-state dynamics saturates the bound identically for time-independent generators, in the time-dependent case, saturation requires a special phase-locking of the drive and is generically lost unless the Hamiltonian commutes with itself at different times.

These relations are exemplified in concrete, physically relevant settings, such as the harmonic oscillator subject to time-dependent frequency dilations and translations, arbitrary closed Virasoro subalgebras, a spin in a rotating magnetic field, and higher-dimensional generalizations for multilevel quantum dynamics~\cite{AllenEberly1975,MorrisShore1983}.

\section{Preliminaries}

\subsection{Krylov subspaces for $\mathfrak{sl}(2,\mathbb{C})$ representations in time-independent dynamics}\label{sec: Time_indep_prelim}

In this subsection, we recall the standard time-independent Krylov construction for operator dynamics~\cite{Viswanath1994,Parker_2019,RabinoviciSanchezGarridoShirSonner2021,Caputa2021_Geometry_KC,HornedalEtAl2022}. Operators are denoted by round kets \(\lvert \mathcal O)\), while ordinary Dirac kets \(\ket{\psi}\) are used for quantum states. The corresponding operator-space Krylov basis is denoted by \(\lvert \mathcal O_n)\). In the Lie-algebraic formalism developed below, the same tridiagonal Krylov structure is written in vectorized Liouville space notation as \(\ket{K_n}\).

The dynamics of an operator generated by a time-independent Liouvillian is
\begin{equation}\label{eq:Op_dyn_time_indep}
    i\partial_t\lvert\mathcal O(t))=\mathcal L\lvert\mathcal O(t)).
\end{equation}
Starting from an initial operator \(\lvert\mathcal O_0)\equiv\lvert\mathcal O(0))\), the Krylov subspace is generated by the set \(\{\mathcal L^n\lvert\mathcal O_0)\}_{n\geq 0}\). Given an inner product, one constructs an orthonormal Krylov basis \(\{\lvert\mathcal O_n)\}\) by the Lanczos recursion
\begin{equation}
    b_{n+1}\lvert\mathcal O_{n+1})
    =\mathcal L\lvert\mathcal O_n)
    -a_n\lvert\mathcal O_n)
    -b_n\lvert\mathcal O_{n-1}),
\end{equation}
with \(\lvert\mathcal O_{-1})=0\), \(b_0=0\), and \((\mathcal O_n\vert\mathcal O_m)=\delta_{nm}\). The recursion terminates when \(b_{d_K}=0\), where \(d_K\) is the Krylov dimension.

The time-independent Lanczos coefficients \(a_n\) and \(b_n\) determine a tridiagonal representation of the Liouvillian in the Krylov basis. In this basis, operator dynamics is mapped to a nearest-neighbor hopping problem on an emergent one-dimensional chain.

Expanding the evolved operator in this basis as
\begin{equation}
    \lvert\mathcal O(t))=\sum_{n=0}^{d_K-1}\varphi_n(t)\lvert\mathcal O_n),
\end{equation}
one finds that the Krylov amplitudes $\varphi_n(t) = (\mathcal{O}_n \lvert \mathcal O(t))$ obey
\begin{equation}
    i\partial_t\varphi_n(t)
    =b_n\varphi_{n-1}(t)+a_n\varphi_n(t)+b_{n+1}\varphi_{n+1}(t),
    \qquad
    \varphi_n(0)=\delta_{n,0}.
\end{equation}
Thus, the dynamics is mapped to the propagation of a particle on a one-dimensional Krylov chain.

For systems with an underlying Lie-algebraic structure, this chain structure is often not merely an abstract numerical construction but can be identified directly with the action of ladder operators in a representation of the corresponding algebra. This observation is especially useful when the generator belongs to a rank-one algebra, as the resulting Krylov chain can be constructed exactly from the ladder structure itself.
In particular, in Ref.~\cite{Caputa2021_Geometry_KC}, it was shown that when the evolution is generated by elements of a Lie algebra, the corresponding Krylov dynamics can be related to generalized coherent-state evolution, thereby providing a geometric description of operator growth; see also Ref.~\cite{HornedalEtAl2022}. This observation motivates the Lie-algebraic viewpoint adopted in the remainder of this work.

Let us now specialize to the case in which the Liouvillian belongs to a representation of
\begin{equation}\label{eq:liu1}
    \mathfrak{sl}(2,\mathbb C)\cong 
    \mathrm{span}_{\mathbb C}\{\mathcal L_+,\mathcal L_-,\mathcal L_0\},
\end{equation}
with commutation relations
\begin{equation}
    [\mathcal L_+,\mathcal L_-]=2\mathcal L_0,
    \qquad
    [\mathcal L_0,\mathcal L_\pm]=\pm \mathcal L_\pm .
\end{equation}
The relevant algebra is selected by the choice of \( * \)-structure. The compact
case corresponds to the \(\mathfrak{su}(2)\) choice
\(\mathcal L_+^\dagger=\mathcal L_-\), while the non-compact case corresponds
to the \(\mathfrak{su}(1,1)\) choice
\(\mathcal L_+^\dagger=-\mathcal L_-\).

For representing Hermitian generators, it is convenient to absorb this sign
into the definition of the lowering operator. We therefore introduce
\begin{equation}
    L_+=\mathcal L_+,\qquad
    L_-=\sigma \mathcal L_-,
    \qquad
    L_0=\mathcal L_0,
\end{equation}
with \(\sigma=+1\) for \(\mathfrak{su}(2)\) and \(\sigma=-1\) for
\(\mathfrak{su}(1,1)\). Then \(L_+^\dagger=L_-\), and the generators obey
\begin{equation}
    [L_+,L_-]=2\sigma L_0,\qquad [L_0,L_\pm]=\pm L_\pm .
\end{equation}
With this convention, both cases can be treated in parallel through the sign
\(\sigma\). If the Liouvillian is of the form
\begin{equation}\label{eq:ladders}
    \mathcal L=\alpha(L_-+L_+),
\end{equation}
then it acts on the basis states as
\begin{equation}\label{eq:ladders2}
    L_+\lvert\mathcal O_n)=b_{n+1}\lvert\mathcal O_{n+1}),\qquad
    L_-\lvert\mathcal O_n)=b_n\lvert\mathcal O_{n-1}).
\end{equation}
 As a result, from Eq.~\eqref{eq:Op_dyn_time_indep} one finds 
\begin{equation}\label{eq:Liouv_time_indep}
    \lvert\mathcal O(t))=e^{-it\alpha(L_++L_-)}\lvert\mathcal O(0)).
\end{equation}
Thus, the Krylov basis is identical to the orbit generated by the ladder operators acting on a lowest-weight state, and the dynamics takes the form of a generalized coherent state.

More generally, whenever the operator Krylov subspace generated by the Liouvillian admits a realization in an ordinary Hilbert-space representation of the same Lie algebra, the vectorized operator basis \(\{\lvert \mathcal O_n)\}\) may be put in one-to-one correspondence with a state basis \(\{\ket{K_n}\}\). Under this identification, the tridiagonal Liouvillian action
\begin{equation}
    \mathcal L\lvert \mathcal O_n)=b_n\lvert \mathcal O_{n-1})+b_{n+1}\lvert \mathcal O_{n+1})
\end{equation}
is represented equivalently as
\begin{equation}
    \mathcal L\ket{K_n}=b_n\ket{K_{n-1}}+b_{n+1}\ket{K_{n+1}},
\end{equation}
where, without notational abuse, we use the same symbol \(\mathcal L\) for the generator in the Liouville 
representation.
In this way, the Krylov structure is preserved, but the discussion can be carried out in a language reminiscent of that for ordinary quantum states.

\subsection{Krylov subspace methods for time-dependent problems}\label{sec:TDK_gen}

We next recall the extension of Krylov subspace methods to explicitly time-dependent generators introduced in Ref.~\cite{takahashi2024_TDKC}. Since the standard Lanczos construction is formulated for a time-independent generator, the basic idea is to embed the Schr\"odinger dynamics into the extended Hilbert space in which time evolution is generated by a time-non-local operator.
For a quantum state obeying the Schr\"odinger equation
\begin{equation}
    i\partial_t \ket{\psi(t)} = H(t)\ket{\psi(t)},
\end{equation}
the dynamics can be embedded into an extended space through the effective generator
\begin{equation}
    H(t)-i\partial_t.
\end{equation}
In this formalism, the time-evolved state may be written as
\begin{equation}
    \ket{\psi(t)}=e^{-it(H(t)-i\partial_t)}\ket{\psi(0)}.
\end{equation}

Treating \(H(t)-i\partial_t\) as the time-dependent analogue of the usual Krylov generator, the Krylov basis \(\{\ket{K_n(t)}\}\) can be constructed starting from the initial state \(\ket{\psi(0)}\) by a generalized Lanczos algorithm. In direct analogy with the time-independent case, the associated coefficients \(a_n(t)\) and \(b_n(t)\) play the role of on-site terms and nearest-neighbor hoppings on the emergent Krylov lattice,
\begin{align}
    b_{n+1}(t)\ket{K_{n+1}(t)}
    &= \left(H(t)-i\partial_t\right)\ket{K_n(t)}
    - a_n(t)\ket{K_n(t)}
    - b_n(t)\ket{K_{n-1}(t)},\\
    a_n(t)
    &= \bra{K_n(t)}\left(H(t)-i\partial_t\right)\ket{K_n(t)},\\
    b_n(t)
    &= \bra{K_{n-1}(t)}\left(H(t)-i\partial_t\right)\ket{K_n(t)}.
\end{align}
The resulting dynamics is therefore mapped to motion on a one-dimensional Krylov lattice with time-dependent nearest-neighbor hopping amplitudes and on-site terms.

Expanding the state on this basis as
\begin{equation}
    \ket{\psi(t)}=\sum_{n=0}^{d_K-1}\varphi_n(t)\ket{K_n(t)},
\end{equation}
the Krylov wavefunction is given by
\begin{equation}
    \varphi_n(t)=\bra{K_n(t)}\ket{\psi(t)},
\end{equation}
and the corresponding Krylov complexity is
\begin{equation}\label{eq:sch}
    K(t)=\sum_{n=0}^{d_K-1} n\,|\varphi_n(t)|^2,
\end{equation}
where \(d_K\) denotes the dimension of the Krylov subspace.

Although this construction provides a conceptually natural extension of the usual Krylov framework, its practical implementation can be challenging. In particular, the explicit determination of the time-dependent Lanczos coefficients and of the associated Krylov wavefunction is typically difficult, and exact results are presently available only in very limited settings~\cite{takahashi2024_TDKC}. One of the main goals of the present work is precisely to identify a broad class of time-dependent Hamiltonians for which this generalized Krylov dynamics can be obtained exactly from the underlying Lie-algebraic structure.

\section{Time-dependent Krylov subspace from Lie algebra}

We now turn to time-dependent Hamiltonians with an underlying Lie-algebraic structure. The required background on the latter can be found in standard expositions \cite{Humphreys1978,DiFrancesco1997}. We provide the general construction under which the Hamiltonian in a properly chosen interaction picture can be written solely in terms of a single pair of ladder operators, and thus its action is tridiagonal in the corresponding lowest-weight representation and directly generates the Krylov basis. We also show below that the exact propagator admits an exponential representation, leading to an effective time-independent Krylov evolution in which the physical time enters only as a fixed parameter.

\subsection{Simple Lie algebras with one-dimensional Krylov lattice}\label{sec:TDK_Lie_gen}

We first consider the case in which the interaction-picture Hamiltonian contains only the ladder operators of an embedded rank-one subalgebra and no Cartan term. The corresponding dynamics is then mapped to nearest-neighbor motion on a one-dimensional Krylov lattice.

Let \(\mathfrak g\) be a simple Lie algebra of rank \(r\), written in Cartan--Weyl form. We denote by \(\mathfrak h\) its Cartan subalgebra, by \(\Delta\) its root system, and by \(\Delta_+\subset\Delta\) the set of positive roots. The algebra is generated by the Cartan elements \(H_i\in\mathfrak h\) and the root operators \(E_\alpha\), \(\alpha\in\Delta\),
\begin{equation}
    \mathfrak g=\mathrm{span}\{H_i,E_\alpha\},
    \qquad i=1,\dots,r,\quad \alpha\in\Delta.
\end{equation}
The Cartan--Weyl commutation relations are
\begin{align}
    [H_i,H_j]&=0,
    & [H_i,E_\alpha]&=\alpha(H_i)E_\alpha, \\
    [E_\alpha,E_{-\alpha}]&=H_\alpha,
    & [H_\alpha,E_{\pm\alpha}]&=\pm 2E_{\pm\alpha},
\end{align}
together with
\begin{equation}
    [E_\alpha,E_\beta]
    =
    \begin{cases}
        N_{\alpha,\beta}\,E_{\alpha+\beta}, & \alpha+\beta\in\Delta,\\
        0, & \alpha+\beta\notin\Delta.
    \end{cases}
\end{equation}
The constants \(N_{\alpha,\beta}\) are the structure constants of the chosen Cartan--Weyl basis; they are nonzero only when \(\alpha+\beta\) is a root, and their precise normalization depends on the normalization convention for the root operators \(E_\alpha\).
Here, \(H_\alpha\) is the coroot associated with \(\alpha\). For simple roots, we choose the Cartan generators as the corresponding simple coroots,
\begin{equation}
    H_i\equiv H_{\alpha_i},
\end{equation}
by which for \(\alpha_j\),
\begin{equation}\label{eq:CartanMatrix}
    [H_i,E_{\alpha_j}]=A_{ji}E_{\alpha_j},
    \qquad
    A_{ji}=\alpha_j(H_i),
\end{equation}
where $A_{ji}$ is the Cartan matrix. In this convention, the row of $A_{ji}$ determines the Cartan action on \(E_{\alpha_j}\), while the embedded rank-one Cartan generator for the \(\alpha_j\) sector is \(H_j/2\).

We consider Hamiltonians of the form
\begin{equation}\label{eq:H_general}
    H(t)=
    \sum_{\alpha\in\Delta_+}\Bigl(f_\alpha(t)E_\alpha+\tilde f_\alpha(t)E_{-\alpha}\Bigr)
    +\sum_{i=1}^{r} g_i(t)H_i,
\end{equation}
where \(\tilde f_\alpha(t)=\sigma_\alpha f_\alpha^*(t)\), with \(\sigma_\alpha=\pm1\), fixes the Hermiticity condition in the corresponding rank-one sector. The choice \(\sigma_\alpha=+1\) gives the compact \(\mathfrak{su}(2)\) case, \(E_\alpha^\dagger=E_{-\alpha}\), while \(\sigma_\alpha=-1\) gives the non-compact \(\mathfrak{su}(1,1)\) case \(E_\alpha^\dagger=-E_{-\alpha}\)~\cite{ban1993decomposition,Zhang1990,Novaes2004}.

We now select a root \(\alpha\in\Delta_+\) and determine the condition for which a suitable interaction-picture transformation yields a Hamiltonian containing only \(E_{\pm\alpha}\). In particular,
\begin{equation}
    L_+=E_\alpha,
    \qquad
    L_-=\sigma_\alpha E_{-\alpha},
    \qquad
    L_0=\frac{1}{2}H_\alpha,
\end{equation}
so that
\begin{equation}
    [L_+,L_-]=2\sigma L_0,
    \qquad
    [L_0,L_\pm]=\pm L_\pm,
\end{equation}
with \(\sigma\equiv\sigma_\alpha\). 
This is the embedded rank-one subalgebra that governs the corresponding Krylov dynamics.
Upon separating the selected ladder pair from the remaining generators, we write
\begin{equation}
    H(t)=f(t)L_+ + f^*(t)L_- + H_E\bigl(\{E_\beta\}_{\beta\neq\alpha},t\bigr)+H_C\bigl(\{H_i\},t\bigr),
\end{equation}
with
\begin{equation}
    H_E\bigl(\{E_\beta\}_{\beta\neq\alpha},t\bigr)
    =
    \sum_{\beta\in\Delta_+,\,\beta\neq\alpha}
    \Bigl(f_\beta(t)E_\beta+\tilde f_\beta(t)E_{-\beta}\Bigr),\qquad
    H_C\bigl(\{H_i\},t\bigr)=\sum_{i=1}^{r} g_i(t)H_i.
\end{equation}

We first remove the Cartan part. Since the \(H_i\) mutually commute, the corresponding interaction-picture transformation is
\begin{equation}
    U_C(t)\coloneqq
    \exp\!\left(-i\sum_{j=1}^r\int_0^t \dd t'\,g_j(t')H_j\right).
\end{equation}
Defining
\begin{equation}\label{eq:phi_beta}
    \phi_\beta(t)\coloneqq
    \sum_{j=1}^r \beta(H_j)\int_0^t \dd t'\,g_j(t'),
\end{equation}
the transformed Hamiltonian becomes
\begin{align}
    H_{I,C}(t)
    &=
    U_C^\dagger(t)
    \left[
        f(t)L_+ + f^*(t)L_- + H_E\bigl(\{E_\beta\}_{\beta\neq\alpha},t\bigr)
    \right]
    U_C(t)
    \notag\\
    &=
    e^{i\phi_\alpha(t)}f(t)L_+
    +
    e^{-i\phi_\alpha(t)}f^*(t)L_-
    +
    H_E^{(I,C)}(t),
\end{align}
where
\begin{equation}
    H_E^{(I,C)}(t)
    \coloneqq
    H_E\bigl(\{e^{i\phi_\beta(t)}E_\beta,e^{-i\phi_\beta(t)}E_{-\beta}\}_{\beta\neq\alpha},t\bigr).
\end{equation}
We now require that the remaining part preserve the chosen rank-one sector,
\begin{equation}
    [H_E^{(I,C)}(t),L_\pm]=[H_E^{(I,C)}(t),L_0]=0.
\end{equation}
A sufficient condition is that, for every root \(\beta\) appearing in \(H_E\), neither \(\beta+\alpha\) nor \(\beta-\alpha\) is a root. In that case, the second interaction-picture transformation generated by \(H_E^{(I,C)}(t)\) leaves the ladder sector unchanged. Defining
\begin{equation}
    U^{(I)}_E(t)\coloneqq
    \mathcal T\exp\!\left[-i\int_0^t \dd t'\,H_E^{(I,C)}(t')\right],
\end{equation}
the final interaction-picture Hamiltonian is
\begin{equation}\label{eq:Ham_Int}
    H_I(t)
    =
    U_E^{(I)\dagger}(t)
    \Bigl(e^{i\phi_\alpha(t)}f(t)L_+
    +
    e^{-i\phi_\alpha(t)}f^*(t)L_-\Bigr)
    U_E^{(I)}(t)
    =
    \gamma(t)L_+ + \gamma^*(t)L_-,
\end{equation}
with
\begin{equation}
    \gamma(t)= e^{i\phi_\alpha(t)}f(t).
\end{equation}
The interaction-picture dynamics is therefore generated entirely by the selected ladder pair.
Applying the exact algorithm of Ref.~\cite{takahashi2024_TDKC} for the lowest-weight state as the initial instantaneous Krylov state as
\begin{eqnarray}
\lvert K_0(t)\rangle&\equiv&\lvert K_0\rangle\equiv\ket{\lambda_1,\ldots,\lambda_r}\equiv\ket{\lambda}\\
    L_-\ket{\lambda}&=&0,
    \qquad
    L_0\ket{\lambda}=\lambda_\alpha\ket{\lambda},
\end{eqnarray}
one finds that the Krylov basis is generated by repeated action of $L_+$ on the lowest-weight state. These states are eigenstates of $L_0$, with $\lambda_\alpha$ the lowest $L_0$-weight in the embedded rank-one sector. For \(H_\alpha=\sum_{i=1}^r c_i H_i\) and \(H_i\ket{\lambda}=\lambda_i\ket{\lambda}\), then
\begin{equation}
    \lambda_\alpha=\frac{1}{2}\sum_{i=1}^{r}c_i\lambda_i.
\end{equation}
One has \(\lambda_\alpha<0\) in the compact \(\mathfrak{su}(2)\) case and \(\lambda_\alpha>0\) in the non-compact \(\mathfrak{su}(1,1)\) case. Since \(U_E^{(I)}(t)\) commutes with \(L_{\pm,0}\) and \(U_C(t)\ket{\lambda}\) differs from \(\ket{\lambda}\) only by a phase factor, the Krylov basis may be constructed directly from \(\ket{\lambda}\) in the final interaction picture.

Using \([L_0,L_+]=L_+\), one finds
\begin{equation}
    L_0L_+^n\ket{\lambda}=(\lambda_\alpha+n)L_+^n\ket{\lambda},
\end{equation}
so the states \(L_+^n\ket{\lambda}\) lie in distinct \(L_0\)-weight spaces. We therefore define
\begin{equation}
    \ket{K_n}
    \coloneqq
    \frac{L_+^n}{\sqrt{\bra{\lambda}L_-^nL_+^n\ket{\lambda}}}\ket{\lambda}.
\end{equation}
The commutation relations give
\begin{equation}
    L_-L_+^n\ket{\lambda}
    =
    -\sigma\,n\bigl(2\lambda_\alpha+n-1\bigr)L_+^{\,n-1}\ket{\lambda},
\end{equation}
and hence
\begin{equation}
    \bra{\lambda}L_-^nL_+^n\ket{\lambda}
    =
    -\sigma\,n\bigl(2\lambda_\alpha+n-1\bigr)
    \bra{\lambda}L_-^{n-1}L_+^{n-1}\ket{\lambda}.
\end{equation}
This fixes the Lanczos coefficients as
\begin{equation}\label{eq:b_n_gen}
    b_n^2
    \coloneqq
    \frac{\bra{\lambda}L_-^nL_+^n\ket{\lambda}}
         {\bra{\lambda}L_-^{n-1}L_+^{n-1}\ket{\lambda}}
    =
    -\sigma\,n\bigl(2\lambda_\alpha+n-1\bigr),
\end{equation}
so that
\begin{equation}
    L_+\ket{K_n}=b_{n+1}\ket{K_{n+1}},
    \qquad
    L_-\ket{K_n}=b_n\ket{K_{n-1}}.
\end{equation}
Accordingly, the interaction-picture Hamiltonian therefore acts tridiagonally,
\begin{equation}\label{eq: H_I_K_n}
    H_I(t)\ket{K_n}
    =
    \gamma(t)b_{n+1}\ket{K_{n+1}}
    +
    \gamma^*(t)b_n\ket{K_{n-1}}.
\end{equation}
We further show by induction that, when the initial state is the lowest-weight state $\ket{K_0}=\ket{\lambda}$, this construction coincides exactly with the time-dependent formulation in Sec.~\ref{sec:TDK_gen}.  First, we show it for the first Krylov state, $\ket{K_1}$. By the time-independence of $\ket{K_0}$ its time-derivative vanishes, $\partial_t\ket{K_0}=0$, and one finds
\begin{eqnarray}
    &&\gamma(t)b_1\ket{K_1(t)}
    = \left(H_I(t)-i\partial_t\right)\ket{K_0}
    - a_0(t)\ket{K_0(t)}= H_I(t)\ket{K_0}
    - a_0(t)\ket{K_0(t)}\nonumber\\
    &&\Rightarrow \ket{K_1(t)}= \frac{L_+\ket{\lambda}}{b_1},
\end{eqnarray}
where the last equation is obtained by normalization and where we assumed the form of $\gamma(t)b_1$ for convenience. As $\ket{K_0}$ is time-independent and $H_I$ contains only the $L_\pm$ ladders, both $\bra{K_0}\partial_t\ket{K_0}$ and $\bra{K_0}L_\pm\ket{K_0}$ vanish, i.e., $a_0=0$. This also identifies the off-diagonal Lanczos coefficient as $b_1=\sqrt{\bra{K_1}\ket{K_1}}$ with the convention as in Eq.~\eqref{eq: H_I_K_n}. Assuming that the construction holds for a given $n$, the next step is obtained by similar arguments,
\begin{eqnarray}
    a_n&=&\bra{K_n}H_I-i\partial_t\ket{K_n}=0,\\
    \gamma(t)b_{n+1}\ket{K_{n+1}}&=&(H_I-i\partial_t)\ket{K_n}-\gamma^*(t)b_{n}\ket{K_{n-1}}=H_I\ket{K_n}-\gamma^*(t)b_{n}\ket{K_{n-1}},\nonumber
\end{eqnarray}
where $a_n=0$ by the time-independence of $\ket{K_n}$ and by $\bra{K_n}L_\pm\ket{K_n}\propto \bra{K_0}L^n_-L_\pm L^n_+\ket{K_0}=0$. As a result, the second line is equivalent to Eq.~\eqref{eq: H_I_K_n} with $b_n$ given in Eq.~\eqref{eq:b_n_gen}.

Finally, we note that the lowest-weight construction directly reproduces the
complexity algebra of Refs.~\cite{HornedalEtAl2022,Caputa2021_Geometry_KC,takahashi2024_TDKC}. Denoting the Krylov Liouvillian as $\hat L=L_++L_-$ and $\hat B=L_+-L_-$, their commutator gives
\begin{eqnarray}
    \hat {\widetilde K}
    \coloneqq [\hat L,\hat B]
    =
    -4\sigma L_0
    =
    -4\sigma(\hat{\mathcal K}+\lambda_\alpha)
    =
    A\hat{\mathcal K}+G,
    \quad
    A=-4\sigma,
    \quad
    G=-4\sigma\lambda_\alpha .
\end{eqnarray}
Here, the Krylov complexity operator is defined as \cite{HornedalEtAl2022}
 \begin{eqnarray}
 \hat{\mathcal K}=\sum_{n}^{d_K-1}n\ket{K_n}\bra{K_n},\quad\hat{\mathcal K}\ket{K_n}=n\ket{K_n},
 \end{eqnarray}
 which is equivalent to the Cartan generator up to a constant shift, $\hat{\mathcal K}=L_0-\lambda_\alpha$.
 The constant $G$ is positive
with the lowest-weight convention: for $\mathfrak{su}(1,1)$ one has
$\sigma=-1$ and $\lambda_\alpha=\kappa>0$, while for $\mathfrak{su}(2)$ one has
$\sigma=+1$ and $\lambda_\alpha=-j$. In the latter finite representation,
$d_K=2j+1$, this gives
$A=-4=-\frac{2G}{d_K-1}$, 
whereas in the infinite-dimensional non-compact representation $A=4\geq0$.
Thus, the shift in $\hat{\mathcal K}=L_0-\lambda_\alpha$ is precisely the constant shift relating the Cartan generator of the embedded rank-one
algebra to the complexity-algebra generator.

\subsection{Krylov-wave function and spread complexity}\label{sec: Krylov_Wavefunction}
Expanding
\begin{equation}\label{eq:l2}
    \ket{\psi_I(t)}
    = U_E^{(I)\dagger}(t)U_C^\dagger(t)\ket{\psi(t)}
    =
    \sum_{n=0}^{d_K-1}\varphi_n(t)\ket{K_n},
    \qquad
    \varphi_n(t)= \bra{K_n}\ket{\psi_I(t)},
\end{equation}
the interaction-picture Schr\"odinger equation 
\begin{equation}
    i\partial_t\ket{\psi_I(t)}=H_I(t)\ket{\psi_I(t)}
\end{equation}
reduces to
\begin{equation}\label{eq:d_t_phi_n}
    i\partial_t\varphi_n(t)
    =
    \gamma(t)b_n\,\varphi_{n-1}(t)
    +
    \gamma^*(t)b_{n+1}\,\varphi_{n+1}(t).
\end{equation}
Thus, the dynamics is mapped to a one-dimensional Krylov lattice with nearest-neighbor hopping only. The corresponding Krylov spread complexity is
\begin{equation}
    K(t)\coloneqq \sum_n n\,|\varphi_n(t)|^2.
\end{equation}
As shown in App.~\ref{app:sch_moving_krylov}, this complexity is invariant under the transformation to the interaction picture, and the Schr\"odinger-picture amplitudes are recovered by the inverse unitary transformation.

The Krylov wavefunction follows from the Wei--Norman disentangling of \(U_I(t)\). Writing
\begin{equation}
    H_I(t)=\gamma(t)L_+ + \gamma^*(t)L_-,
\end{equation}
the corresponding evolution operator admits the Wei--Norman factorization
\begin{equation}
    U_I(t)\coloneqq \mathcal T\exp\!\left[-i\int_0^t\dd t'\,H_I(t')\right]
    =
    e^{z(t)L_+}e^{\eta(t)L_0}e^{w(t)L_-},
\end{equation}
where the Wei--Norman coefficients satisfy~\cite{Charzynski_2013}
\begin{eqnarray}\label{eq:Wei_Norman_eqs}
    \dot z&=&-i\gamma(t)+i\sigma \gamma^*(t)z^2(t),\qquad
    \dot\eta(t)=2i\sigma \gamma^*(t)z(t),\\
    \dot w(t)&=&-i\gamma^*(t)e^{\eta(t)},\qquad \qquad \qquad z(0)=\eta(0)=w(0)=0.
\end{eqnarray}
For the lowest-weight initial state
\begin{equation}
    \ket{\psi_I(0)} = \ket{\lambda} \equiv \ket{0},
    \qquad
    L_-\ket{0}=0,
    \qquad
    L_0\ket{0}=\lambda_\alpha\ket{0},
\end{equation}
one obtains
\begin{equation}
    \ket{\psi_I(t)}
    =
    U_I(t)\ket{0}
    =
    e^{z(t)L_+}e^{\eta(t)L_0}\ket{0}
    =
    e^{\lambda_\alpha\eta(t)}e^{z(t)L_+}\ket{0}.
\end{equation}
Expanding the exponential gives
\begin{equation}
    \ket{\psi_I(t)}
    =
    e^{\lambda_\alpha\eta(t)}
    \sum_{n=0}^{d_K-1}\frac{z^n(t)}{n!}L_+^n\ket{0}
    =
    e^{\lambda_\alpha\eta(t)}
    \sum_{n=0}^{d_K-1}
    z^n(t)\,
    \frac{\sqrt{\bra{0}L_-^nL_+^n\ket{0}}}{n!}\,
    \ket{K_n}.
\end{equation}

For \(\sigma=1\), the algebra is compact and \(d_K=2j+1\), with \(\lambda_\alpha=-j\). This yields
\begin{eqnarray}
    L_+^n\lvert j,-j\rangle&\propto&\lvert j,n-j\rangle,\quad n=0,1,\dots, 2j,\\
    \bra{0}L_-^nL_+^n\ket{0}
    &=&
    \prod_{m=1}^{n}m(2j-m+1)
    =
    n!\,\frac{(2j)!}{(2j-n)!},\\
    \frac{\sqrt{\bra{0}L_-^nL_+^n\ket{0}}}{n!}
    &=&
    \sqrt{\frac{(2j)!}{n!\,(2j-n)!}}
    =
    \sqrt{\binom{2j}{n}}.
\end{eqnarray}
Therefore,
\begin{eqnarray}\label{eq:varphi_n_sol}
    \ket{\psi_I(t)}
    &=&
    e^{-j\eta(t)}
    \sum_{n=0}^{2j}
    z^n(t)\sqrt{\binom{2j}{n}}\,\ket{K_n},\\
    \varphi_n(t)&=&e^{-j\eta(t)}z^n(t)\sqrt{\binom{2j}{n}},
\end{eqnarray}
with $n=0,1,\dots,2j$. 
Using unitarity, \(e^{-\eta(t)-\eta^*(t)}=\bigl(1+|z(t)|^2\bigr)^{-2}\), and so the Krylov probabilities depend only on \(z(t)\),
\begin{equation}\label{eq:P_n_su_2}
    P_n(t)\coloneqq |\varphi_n(t)|^2
    =
    \binom{2j}{n}\,
    \frac{|z(t)|^{2n}}{\bigl(1+|z(t)|^2\bigr)^{2j}},
    \qquad n=0,1,\dots,2j.
\end{equation}
The complexity is then
\begin{equation}
    K(t)\coloneqq \sum_{n=0}^{2j}n\,P_n(t)
    =
    \frac{2j\,|z(t)|^2}{1+|z(t)|^2}
    =
    \frac{-2\lambda_\alpha\,|z(t)|^2}{1+|z(t)|^2}.
\end{equation}

For \(\sigma=-1\), writing \(\kappa=\lambda_\alpha>0\), one has \(b_n^2=n(2\kappa+n-1)\) and
\begin{equation}
    \bra{0}L_-^nL_+^n\ket{0}
    =
    \prod_{m=1}^{n}m(2\kappa+m-1)
    =
    n!\,\frac{\Gamma(2\kappa+n)}{\Gamma(2\kappa)},\quad
    \frac{\sqrt{\bra{0}L_-^nL_+^n\ket{0}}}{n!}
    =
    \sqrt{\frac{\Gamma(2\kappa+n)}{n!\,\Gamma(2\kappa)}}.
\end{equation}
Hence,
\begin{eqnarray}
    \ket{\psi_I(t)}
    &=&
    e^{\kappa\eta(t)}
    \sum_{n=0}^{\infty}
    z^n(t)\sqrt{\frac{\Gamma(2\kappa+n)}{n!\,\Gamma(2\kappa)}}\,\ket{K_n},\\    
    \varphi_n(t)&=&e^{\kappa\eta(t)}z^n(t)\sqrt{\frac{\Gamma(2\kappa+n)}{n!\,\Gamma(2\kappa)}},
    \qquad n=0,1,2,\dots,
\end{eqnarray}
and by unitarity
\begin{equation}
    e^{\eta(t)+\eta^*(t)}=\bigl(1-|z(t)|^2\bigr)^2,
    \qquad |z(t)|<1.
\end{equation}
The Krylov occupation probabilities become
\begin{equation}\label{eq:P_n_su_11}
    P_n(t)\coloneqq |\varphi_n(t)|^2
    =
    \frac{\Gamma(2\kappa+n)}{n!\,\Gamma(2\kappa)}
    \bigl(1-|z(t)|^2\bigr)^{2\kappa}|z(t)|^{2n},
    \qquad n=0,1,2,\dots
\end{equation}
which is a negative-binomial distribution, yielding the following expression for the Krylov complexity:
\begin{equation}
    K(t)\coloneqq \sum_{n=0}^{\infty}n\,P_n(t)
    =
    \frac{2\kappa\,|z(t)|^2}{1-|z(t)|^2}
    =
    \frac{2\kappa\,|z(t)|^2}{1-|z(t)|^2}.
\end{equation}
In a unified form, the complexity reads
\begin{equation}
    K(t)=\frac{-2\sigma\,\lambda_\alpha\,|z(t)|^2}{1+\sigma |z(t)|^2},
\end{equation}
where \(z(t)\) denotes the solution of Eq.~\eqref{eq:Wei_Norman_eqs} for the given \(\sigma=\pm1\).

Finally, we compare this construction with the standard Krylov approach based on the exponential expression of the time-evolution operator,
\begin{equation}\label{eq: U_G}
    U_I(t)=e^{-iG(t)},
    \qquad
    G(t)=\Theta_0(t)\,L_0+\Theta_+(t)\,L_+ + \Theta_+^*(t)\,L_-,
    \qquad
    \Theta_0\in\mathbb{R},
\end{equation}
where \(G(t)\) remains in the same rank-one algebra. The parameters \((\Theta_0,\Theta_+)\) are determined exactly from the Wei--Norman variables \((z,\eta)\) by the inversion formulas derived in App.~\ref{app:U_G}. Defining \(A\coloneqq e^{-\eta/2}\) and \(B\coloneqq z\,e^{-\eta/2}\), one finds
\begin{equation}\label{eq:Theta_inversion_main}
    \Theta_0
    =\frac{2\chi\,\operatorname{Im}A}
          {\sqrt{(\operatorname{Im}A)^2+\sigma|B|^2}}\,,
    \quad
    \Theta_+
    =\frac{i\chi\,B}
          {\sqrt{(\operatorname{Im}A)^2+\sigma|B|^2}}\,,
    \quad
    \chi
    =\arctan\frac{\sqrt{(\operatorname{Im}A)^2+\sigma|B|^2}}
                  {\operatorname{Re}A}\,,
\end{equation}
with the \(\arctan\) branch fixed by continuity from \(\chi(0)=0\).
For $\Theta_+(t)\neq0$, this generator $G(t)$ induces the same ladder Krylov basis ${\ket{K_n}}$,
\begin{eqnarray}
L_0\ket{K_n}&=&(\lambda_\alpha+n)\ket{K_n},\quad
L_+\ket{K_n}=b_{n+1}\ket{K_{n+1}},\quad
L_-\ket{K_n}=b_n\ket{K_{n-1}},\\
    G(t)\ket{K_n}
    &=&
    a_n(t)\ket{K_n}
    +
    \tilde b_{n+1}(t)\ket{K_{n+1}}
    +
    \tilde b_n^*(t)\ket{K_{n-1}},
\end{eqnarray}
with
\begin{equation}
    a_n(t)=\Theta_0(t)(\lambda_\alpha+n),
    \qquad
    \tilde b_n(t)=\Theta_+(t)\,b_n,
    \qquad
    b_n=\sqrt{-\sigma\,n\bigl(2\lambda_\alpha+n-1\bigr)}.
\end{equation}
If $\Theta_+(t)=0$, the auxiliary problem is diagonal and the Krylov dynamics is trivial.

This complementary construction differs from the interaction-picture dynamics~\eqref{eq:d_t_phi_n} in two ways: it has a diagonal term \(a_n(t)\), and its off-diagonal coefficients carry the phase of \(\Theta_+(t)\). More importantly, it is time-independent in an auxiliary parameter \(s\), while the physical time \(t\) enters only parametrically through \(a_n(t)\) and \(\tilde b_n(t)\). The corresponding amplitudes satisfy
\begin{equation}
    i\partial_s\psi_n(s)
    =
    a_n\,\psi_n(s)
    +
    \tilde b_{n+1}\,\psi_{n+1}(s)
    +
    \tilde b_n^*\,\psi_{n-1}(s),
    \qquad
    \psi_n(0)=\delta_{n,0},
\end{equation}
with solution
\begin{equation}
    \psi_n(s)= \bra{K_n}e^{-is G}\ket{0}.
\end{equation}
Since \(e^{-iG(t)}=U_I(t)\), evaluating at \(s=1\) gives
\begin{equation}
    \psi_n(1)=\braket{K_n}{U_I(t)|0}=\varphi_n(t),
\end{equation}
so the effective time-independent evolution reproduces the exact physical Krylov wavefunction at \(s=1\).

The evolution remains purely off-diagonal only when \(\gamma(t)=e^{i\delta}r(t)\) with constant phase \(\delta\). In that case, time ordering is trivial, \(\Theta_0=0\), and
\begin{equation}
    \Theta_+=\int_0^t\!\mathrm{d}s\,\gamma(s)
    =
    e^{i\delta}\!\int_0^t\!\mathrm{d}s\,r(s),
\end{equation}
so that
\begin{equation}
    G(t)=\left(e^{i\delta}L_++e^{-i\delta}L_-\right)
    \int_0^t\!\mathrm{d}s\,r(s).
\end{equation}
Then \(a_n=0\) and \(\tilde b_n=b_n\int_0^t\!\mathrm{d}s\,\gamma(s)\). The generator thus contains the time-integrated hopping amplitudes of the physical Krylov dynamics~\eqref{eq:d_t_phi_n}. For a time-dependent phase of \(\gamma(t)\), one generically has \(\Theta_0\neq0\), and the diagonal term cannot be removed without reintroducing a time-dependent phase in the hopping; see App.~\ref{app:U_G}.

\section{Demonstration on exactly solvable models}
In this section, we provide examples of the general Lie-algebraic framework in exactly solvable models.

\subsection{Multi-level systems with $\mathfrak{su}(4)$}

To illustrate the general construction, we consider the Lie algebra \(\mathfrak{su}(4)\), which provides one of the simplest examples containing two commuting simple-root sectors. Since \(\mathfrak{su}(4)\) is of type \(A_3\) with rank \(r=3\), its Dynkin diagram is
\begin{equation*}
\begin{tikzpicture}[scale=1.5, baseline=-0.5ex]
    \foreach \x in {1,2,3}
        \draw[fill=white] (\x,0) circle (.1) node[below=3pt] {$\alpha_{\x}$};
    \draw (1.1,0) -- (1.9,0);
    \draw (2.1,0) -- (2.9,0);
\end{tikzpicture}
\end{equation*}
and the full set of positive roots is
\begin{equation}
\Delta_+=\{\alpha_1,\alpha_2,\alpha_3,\alpha_1+\alpha_2,\alpha_2+\alpha_3,\alpha_1+\alpha_2+\alpha_3\}.
\end{equation}

The Cartan matrix and the corresponding commutation relations between the Cartan elements \(H_k\), \(k=1,2,3\), and the simple-root generators are
\begin{equation}
A_{jk}=
\begin{pmatrix}
2 & -1 & 0\\
-1 & 2 & -1\\
0 & -1 & 2
\end{pmatrix},
\qquad
[H_k,H_\ell]=0,
\qquad
[H_k,E_{\alpha_j}]=A_{jk}E_{\alpha_j}.
\end{equation}
Equivalently, the rows of the Cartan matrix encode the Cartan action on the simple-root generators,
\begin{equation}
    \alpha_1(H_k)=(2,-1,0),
    \qquad
    \alpha_2(H_k)=(-1,2,-1),
    \qquad
    \alpha_3(H_k)=(0,-1,2).
\end{equation}
Each simple-root pair closes an \(\mathfrak{su}(2)\) subalgebra,
\begin{equation}
    [E_{\alpha_i},E_{-\alpha_i}]=H_i,
    \qquad i=1,2,3,
\end{equation}
where \(H_i\equiv H_{\alpha_i}\) are the simple coroots. Thus, the rows of \(A_{jk}\) determine the Cartan phases of the root generators, while the embedded rank-one Cartan generator for the \(\alpha_i\) sector is \(H_i/2\).
Among the positive-root generators, the only non-vanishing commutators are
\begin{eqnarray}
& & [E_{\alpha_1},E_{\alpha_2}]\propto E_{\alpha_1+\alpha_2},\\
& & [E_{\alpha_2},E_{\alpha_3}]\propto E_{\alpha_2+\alpha_3},\\
& & [E_{\alpha_1},E_{\alpha_2+\alpha_3}]\propto E_{\alpha_1+\alpha_2+\alpha_3},\\
& & [E_{\alpha_1+\alpha_2},E_{\alpha_3}]\propto E_{\alpha_1+\alpha_2+\alpha_3}.
\end{eqnarray}
In particular, since neither \(\alpha_1+\alpha_3\) nor \(\alpha_1-\alpha_3\) is a root, the simple-root sectors generated by \(\alpha_1\) and \(\alpha_3\) commute.

We now apply the general construction with \(L_\pm\coloneqq E_{\pm\alpha_1},\,L_0=H_1/2\), and consider the Hamiltonian
\begin{equation}
    H(t)=f(t)L_+ + f^*(t)L_- + f_{\alpha_3}(t)E_{\alpha_3}+f_{\alpha_3}^*(t)E_{-\alpha_3}+\sum_{i=1}^3 g_i(t)H_i.
\end{equation}
Because the \(\alpha_3\) sector commutes with the embedded \(\alpha_1\) subalgebra, it factors from the \(\alpha_1\) dynamics into the commuting operator \(U_E^{(I)}(t)\). The interaction-picture Hamiltonian governing the Krylov dynamics in the \(\alpha_1\) sector is therefore
\begin{equation}
    H_I^{(\alpha_1)}(t)=\gamma(t)L_+ + \gamma^*(t)L_-,
    \qquad
    \gamma(t)=f(t)\exp\!\left(i\int_0^t\dd t'\,[2g_1(t')-g_2(t')]\right).
\end{equation}
From a physical perspective, this choice is also natural in view of reductions of multilevel driven dynamics to independent effective two-state sectors in the sense of Morris--Shore~\cite{MorrisShore1983}, together with hyperbolic-secant pulses familiar from exactly solvable two-level driving protocols such as the Allen--Eberly and Rosen--Zener schemes~\cite{AllenEberly1975}.
A convenient choice that renders \(\gamma(t)\) real and simultaneously keeps the commuting \(\alpha_3\) sector analytically tractable is
\begin{equation}
    f(t)=f_{\alpha_3}(t)=\Omega_0\,\sech\!\left(\frac{t}{T}\right),\qquad
    g_1(t)=g_3(t)=\Delta(t),\qquad
    g_2(t)=2\Delta(t),
\end{equation}
with arbitrary real \(\Delta(t)\). For this choice, the phases from the Cartan parts cancel, so that
\begin{equation}
    \gamma(t)=f(t)=\Omega_0\,\sech\!\left(\frac{t}{T}\right).
\end{equation}
The \(z(t)\) solution of the Wei--Norman equations in Eq.~\eqref{eq:Wei_Norman_eqs} with \(\sigma=1\) acquires the exact form
\begin{equation}
    z(t)=-i\tan\Theta(t),\qquad
    \eta(t)=-2\ln\!\bigl[\cos\Theta(t)\bigr],\qquad
    w(t)=-i\tan\Theta(t),
\end{equation}
where
\begin{equation}
    \Theta(t)\coloneqq\int_0^t\dd t'\,\gamma(t')
    =
    2\Omega_0T\,\arctan\!\left[\tanh\!\left(\frac{t}{2T}\right)\right].
\end{equation}

For the simplest non-trivial representation of the \(\alpha_1\) subalgebra, the lowest weight is \(\lambda_{\alpha_1}=-1\), corresponding to \(j=1\). The Krylov space is then three-dimensional, with \(n=0,1,2\). The Lanczos coefficients follow from the general expression, Eq.~\eqref{eq:b_n_gen}, as
\begin{equation}
    b_n=\sqrt{n(3-n)},
    \qquad n=0,1,2.
\end{equation}
The Krylov wavefunction is obtained from the general \(\mathfrak{su}(2)\) result~\eqref{eq:varphi_n_sol},
\begin{equation}
    \varphi_n(t)=(-i)^n\sqrt{\binom{2}{n}}\,
    \tan^n\Theta(t)\,\cos^2\Theta(t),
    \qquad n=0,1,2,
\end{equation}
and the Krylov complexity is
\begin{equation}
    K(t)=2\sin^2\Theta(t)
    =
    2\sin^2\!\left(
    2\Omega_0T\,\arctan\!\left[\tanh\!\left(\frac{t}{2T}\right)\right]
    \right).
\end{equation}
\begin{figure}[t]
\centering
\includegraphics[width=.47\linewidth,trim={2.5cm 0 7cm 0},clip]{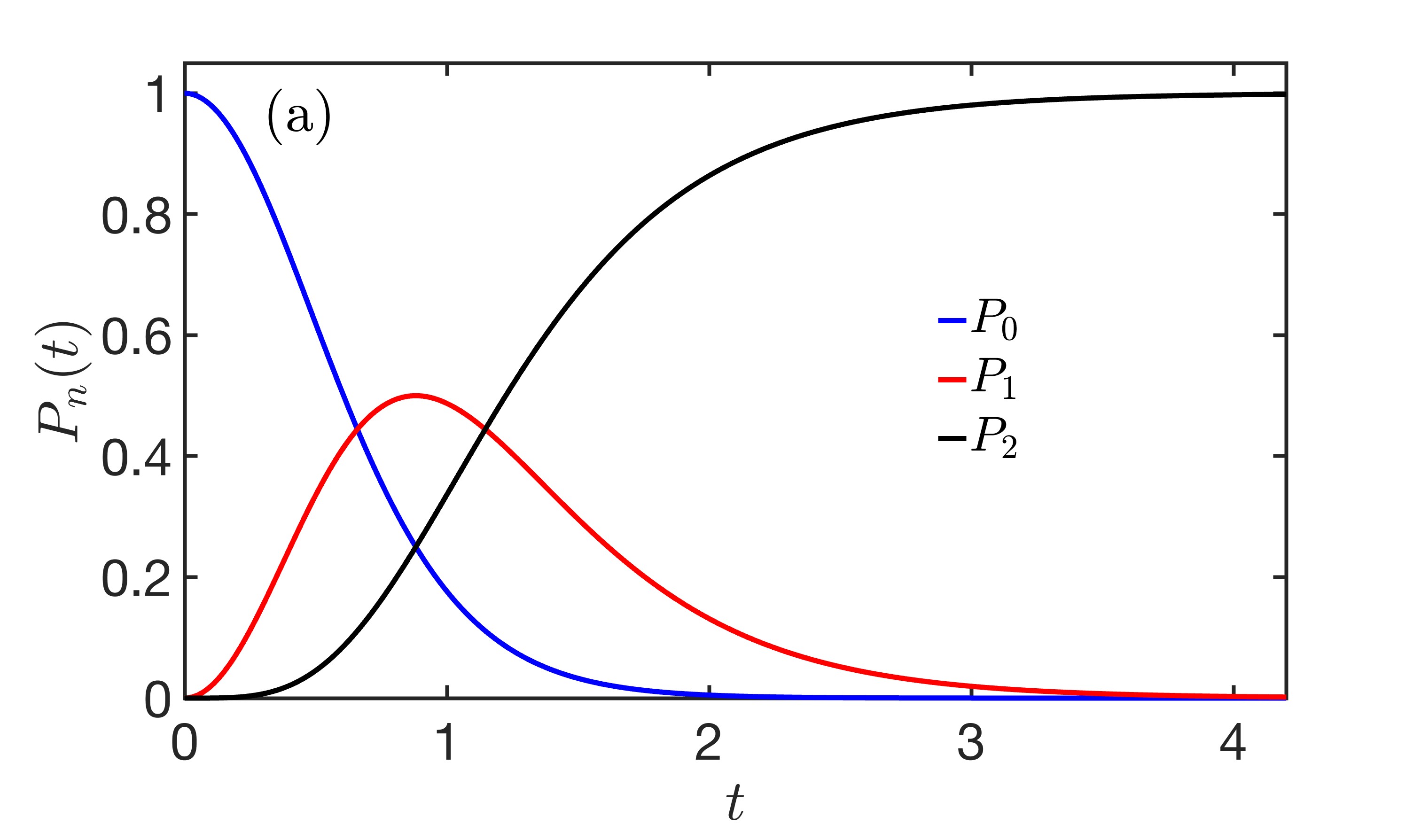}
\includegraphics[width=.47\linewidth,trim={2.5cm 0 7cm 0},clip]{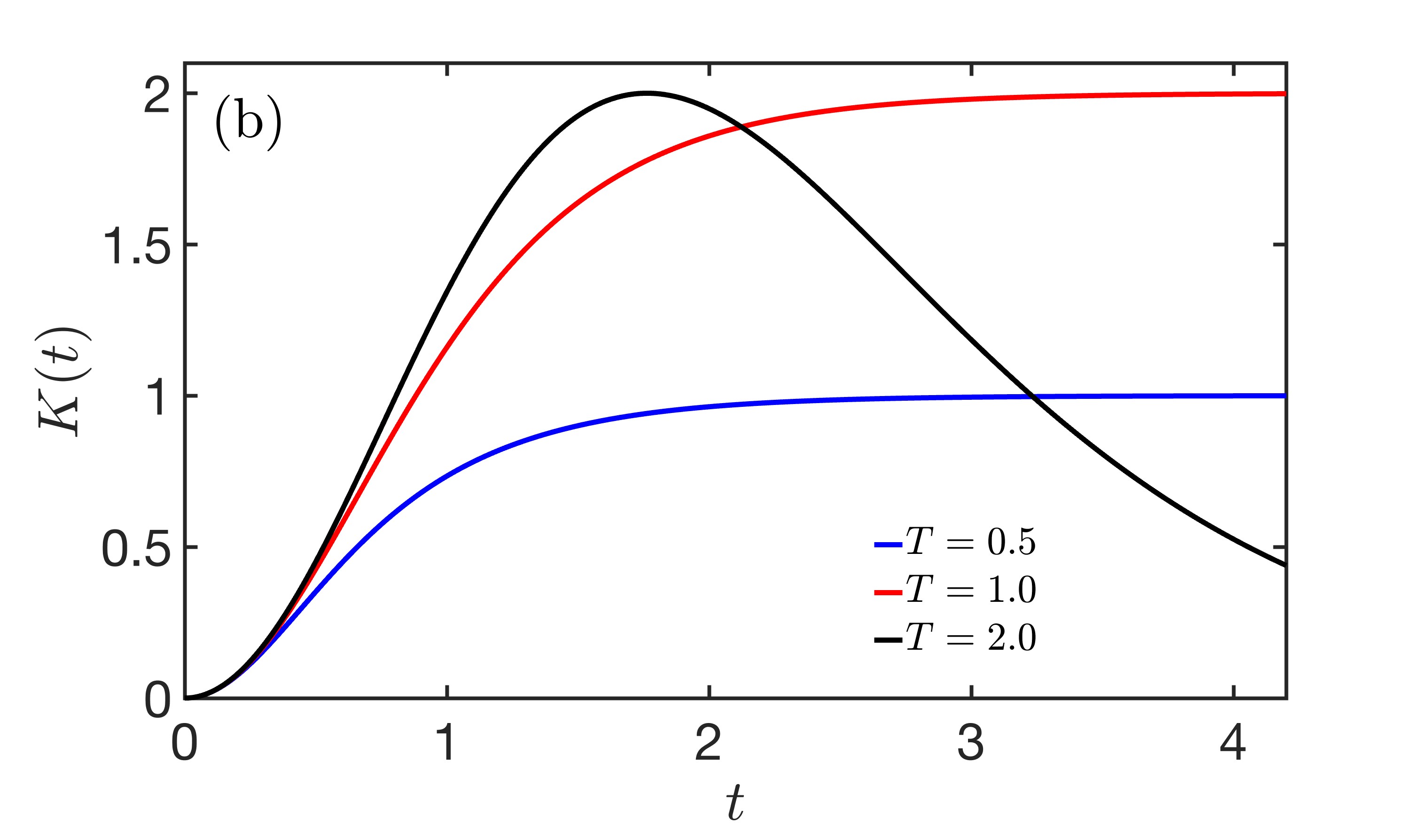}
\caption{(a) Krylov occupation probabilities in the \(\mathfrak{su}(4)\) example for \(T=1\) and \(\Omega_0=1\), showing transfer of weight to the highest Krylov state at long times, with \(P_2\to1\). (b) Krylov complexity for \(T=0.5,1,2\) and \(\Omega_0=1\), approaching the corresponding long-time values \(1\), \(2\), and \(0\), respectively.}
\label{fig:su_4}
\end{figure}

The corresponding occupation probabilities, \(P_n(t)=|\varphi_n(t)|^2\), are shown in Fig.~\ref{fig:su_4}a for \(T=1\). At long times, the probability weight is transferred to the highest Krylov state, consistent with \(P_2\to1\). The Krylov complexity is displayed in Fig.~\ref{fig:su_4}b for \(T=0.5,1,2\). The asymptotic value is controlled by the pulse area $\Theta(\infty)=\pi\Omega_0 T/2$, leading to the corresponding final complexities shown in Fig.~\ref{fig:su_4}b as well.

\subsection{Harmonic oscillator with dilation}

We next consider the dilated harmonic oscillator~\cite{TORRONTEGUI2013_STA,Lohe_2009_HoExact,Muga_2010_Ho_STA,coelho2024TimeDepHO_LR}, which provides a standard realization of the non-compact case,
\begin{equation}
    H(t)=\frac{p^2}{2m}+\frac{1}{2}m\omega^2(t)\,x^2
    =g(t)\,a^\dagger a+f(t)\bigl[a^2+(a^\dagger)^2\bigr]+E_0(t),
\end{equation}
where
\begin{equation}
    a=\sqrt{\frac{m\omega_0}{2}}\!\left(x+\frac{i}{m\omega_0}p\right),
    \quad
    g(t)=\frac{\omega^2(t)+\omega_0^2}{2\omega_0},
    \quad
    f(t)=\frac{\omega^2(t)-\omega_0^2}{4\omega_0},
    \quad
    E_0(t)=\frac{g(t)}{2},
\end{equation}
with \(\omega_0\equiv\omega(0)\). The squeezing sector is governed by an embedded \(\mathfrak{su}(1,1)\) subalgebra generated by
\begin{equation}
    L_+=\tfrac{1}{2}(a^\dagger)^2,
    \qquad
    L_-=\tfrac{1}{2}a^2,
    \qquad
    L_0=\tfrac{1}{4}(2a^\dagger a+1),
\end{equation}
satisfying \([L_+,L_-]=-2L_0\), \([L_0,L_\pm]=\pm L_\pm\), i.e.,\ \(\sigma=-1\). Without loss of generality, we choose the lowest-weight state to be the vacuum annihilated by \(a\). Hence, \(L_-\ket{0}=0\) and \(L_0\ket{0}=\frac{1}{4}\ket{0}\), so that \(\lambda_\alpha=\frac{1}{4}\).
Passing to the interaction picture with respect to \(H_0(t)=g(t)(a^\dagger a+\frac{1}{2})\) and discarding the scalar \(E_0(t)\), one obtains
\begin{equation}
    H_I(t)
    =
    \gamma(t)\,L_+ + \gamma^*(t)\,L_-,
    \qquad
    \gamma(t)=2f(t)\,e^{2i\mathcal G(t)},
    \qquad
    \mathcal G(t)=\int_0^t\!\mathrm{d}t'\,g(t').
\end{equation}
The Krylov basis and the Lanczos coefficients are
\begin{equation}
    \ket{K_n} =
    \frac{(a^\dagger)^{2n}}{\sqrt{\bra{0}a^{2n}(a^\dagger)^{2n}\ket{0}}}\ket{0}
    = \ket{2n}, \qquad
    b_n = \sqrt{\frac{n(2n-1)}{2}},
\end{equation}
respectively, where \(\ket{2n}\) denotes the even-parity harmonic-oscillator eigenstates. The Krylov wavefunction obeys
\begin{eqnarray}
i\partial_t\varphi_n(t)
    &=&
    \gamma(t)\,b_n\,\varphi_{n-1}(t)
    +
    \gamma^*(t)\,b_{n+1}\,\varphi_{n+1}(t),
\end{eqnarray}
and the Wei--Norman equations~\eqref{eq:Wei_Norman_eqs} with \(\sigma=-1\) read explicitly
\begin{equation}\label{eq:WN_su11_dilated}
    \dot z=-i\gamma(t)-i\gamma^*(t)\,z^2,
    \qquad
    \dot\eta=-2i\gamma^*(t)\,z,
    \qquad
    z(0)=\eta(0)=0.
\end{equation}
From the general \(\mathfrak{su}(1,1)\) solution with \(\lambda_\alpha=\frac{1}{4}\), the Krylov probabilities and complexity are
\begin{eqnarray}
    P_n(t)
   &=&
    \frac{(2n)!}{4^n(n!)^2}
    \,\bigl(1-|z(t)|^2\bigr)^{1/2}\,|z(t)|^{2n},\\    
    K(t)
    &=&
    \frac{|z(t)|^2}{2\bigl(1-|z(t)|^2\bigr)},
\end{eqnarray}
respectively.

As an explicit example, we consider the sudden frequency quench~\cite{Coelho_2022}
\begin{equation}\label{eq:freq_quench}
    \omega(t)=
    \begin{cases}
        \omega_0, & t<0,\\
        \omega_1, & 0<t<\tau,\\
        \omega_0, & t>\tau,
    \end{cases}
\end{equation}
for which \(f\) and \(g\) are piecewise constant. During the quench (\(0<t<\tau\)), the coupling reads \(\gamma(t)=2f_0\,e^{2ig_0 t}\) with
\begin{equation}
    f_0=\frac{\omega_1^2-\omega_0^2}{4\omega_0},
    \qquad
    g_0=\frac{\omega_1^2+\omega_0^2}{2\omega_0}\,.
\end{equation}
Since \(|\gamma|=2|f_0|\) is constant but the phase rotates at rate \(2g_0\), the Wei--Norman equations~\eqref{eq:WN_su11_dilated} are reduced to constant coefficients by the substitution \(z(t)=e^{2ig_0 t}\zeta(t)\), giving
\begin{equation}
    \dot\zeta=-2if_0(1+\zeta^2)-2ig_0\,\zeta,
    \qquad
    \zeta(0)=0.
\end{equation}
Linearizing via \(\zeta=\dot u/(2if_0\,u)\) yields \(\ddot u+2ig_0\,\dot u-4f_0^2\,u=0\) with characteristic roots \(\mu=i(-g_0\pm\omega_1)\), where we used the identity \(4f_0^2-g_0^2=-\omega_1^2\). The solution satisfying \(u(0)=1\), \(\dot u(0)=0\) is
\begin{equation}
    u(t)=e^{-ig_0 t}\!\left[\cos\omega_1 t
    +i\frac{g_0}{\omega_1}\sin\omega_1 t\right],
\end{equation}
from which we obtain
\begin{equation}
    z(t)
    =
    \frac{-2if_0\,e^{2ig_0 t}\sin\omega_1 t}
    {\omega_1\cos\omega_1 t+ig_0\sin\omega_1 t}\,,
    \qquad
    |z(t)|^2
    =
    \frac{4f_0^2\sin^2\omega_1 t}
    {\omega_1^2+4f_0^2\sin^2\omega_1 t}\,.
\end{equation}
The equation involving \(\eta\) integrates to \(\eta=-2\ln u\), giving
\begin{equation}
    \eta(t)
    =
    2ig_0\,t
    -2\ln\!\left[\cos\omega_1 t
    +i\frac{g_0}{\omega_1}\sin\omega_1 t\right].
\end{equation}

\begin{figure}[t]
\centering
\includegraphics[width=.47\linewidth,trim={1.5cm 0 7cm 0},clip]{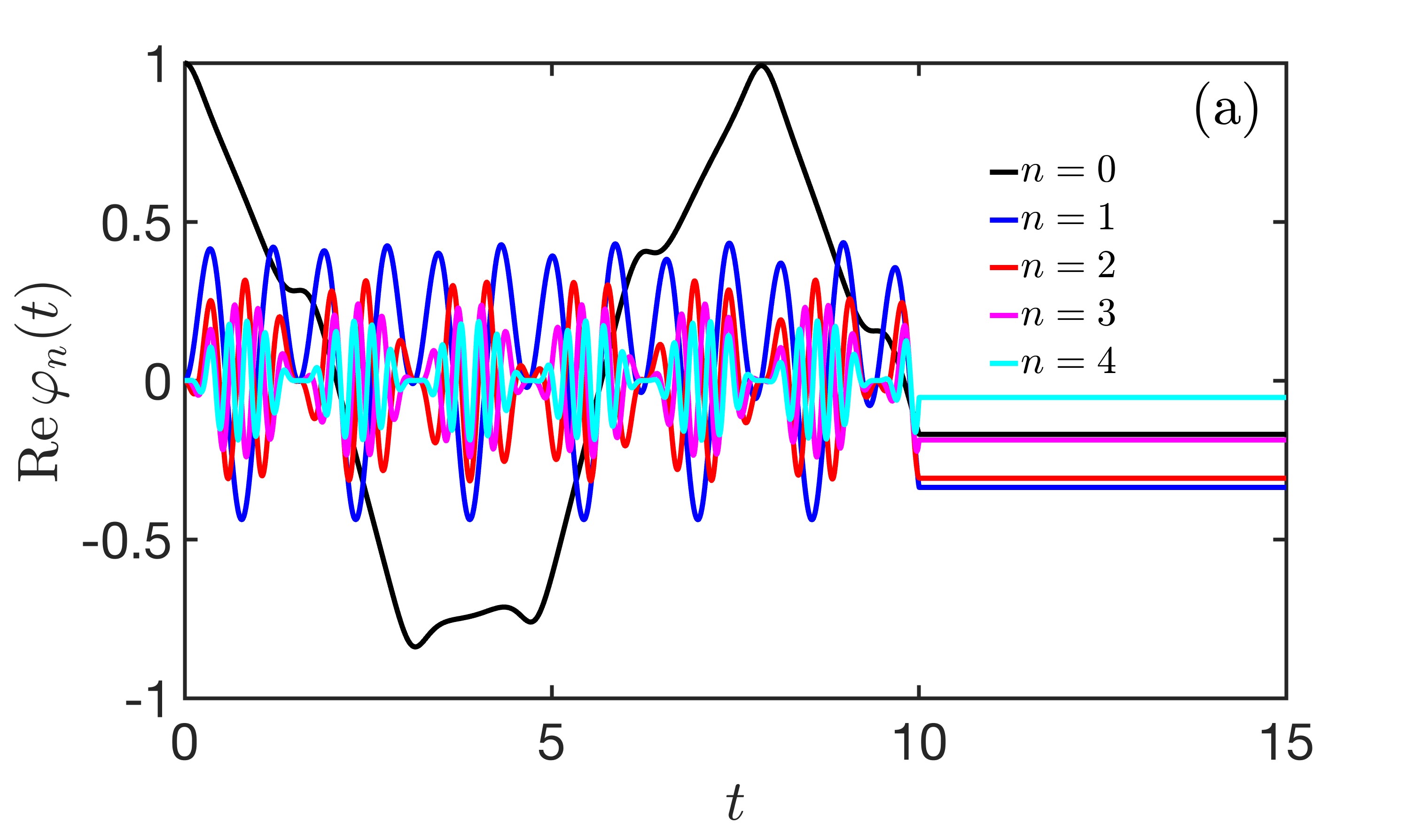}
\includegraphics[width=.47\linewidth,trim={1.5cm 0 7cm 0},clip]{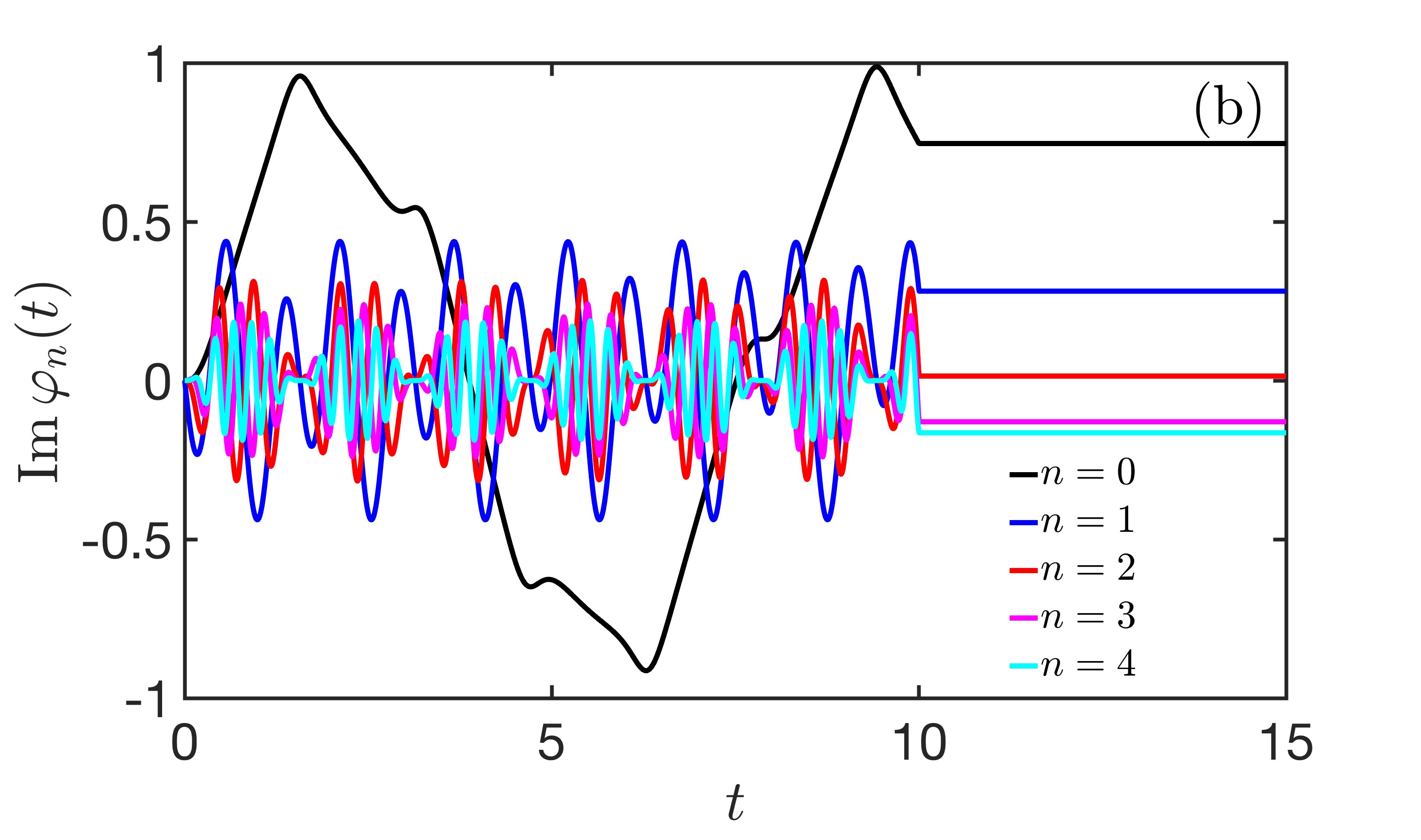}
\caption{(a) Real part of the Krylov wavefunction for a representative frequency quench, oscillating in the interval \(0<t<\tau\) and freezing to a constant value for \(t>\tau\). (b) Imaginary part of the Krylov wavefunction, showing the same qualitative behaviour. ($\omega_0=0.6,\,\omega_1=2,\,\tau=10$)}
\label{fig:HO_dil_phi}
\end{figure}

\begin{figure}[t]
\centering
\includegraphics[width=.47\linewidth,trim={2.5cm 0 7cm 0},clip]{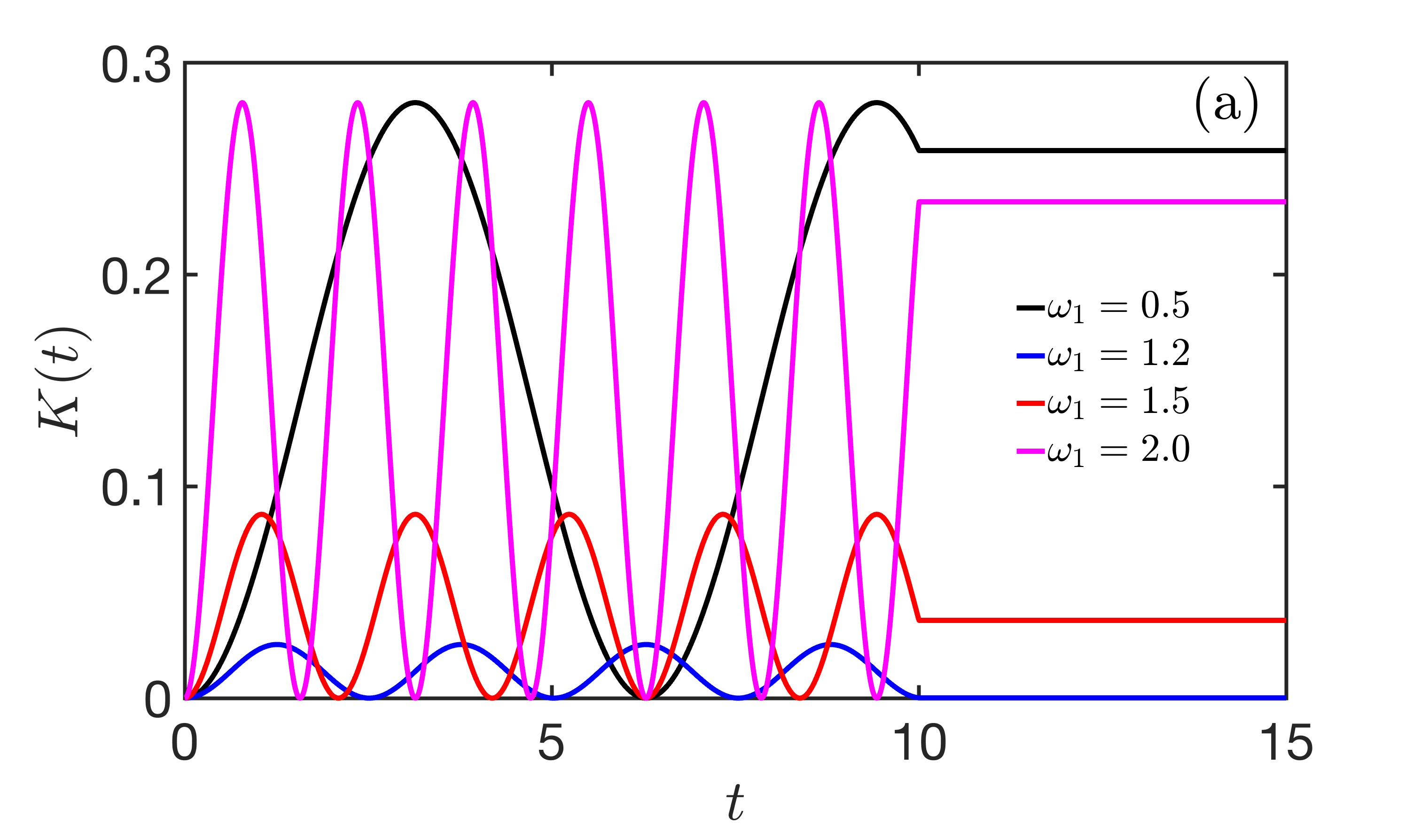}
\includegraphics[width=.47\linewidth,trim={2.5cm 0 7cm 0},clip]{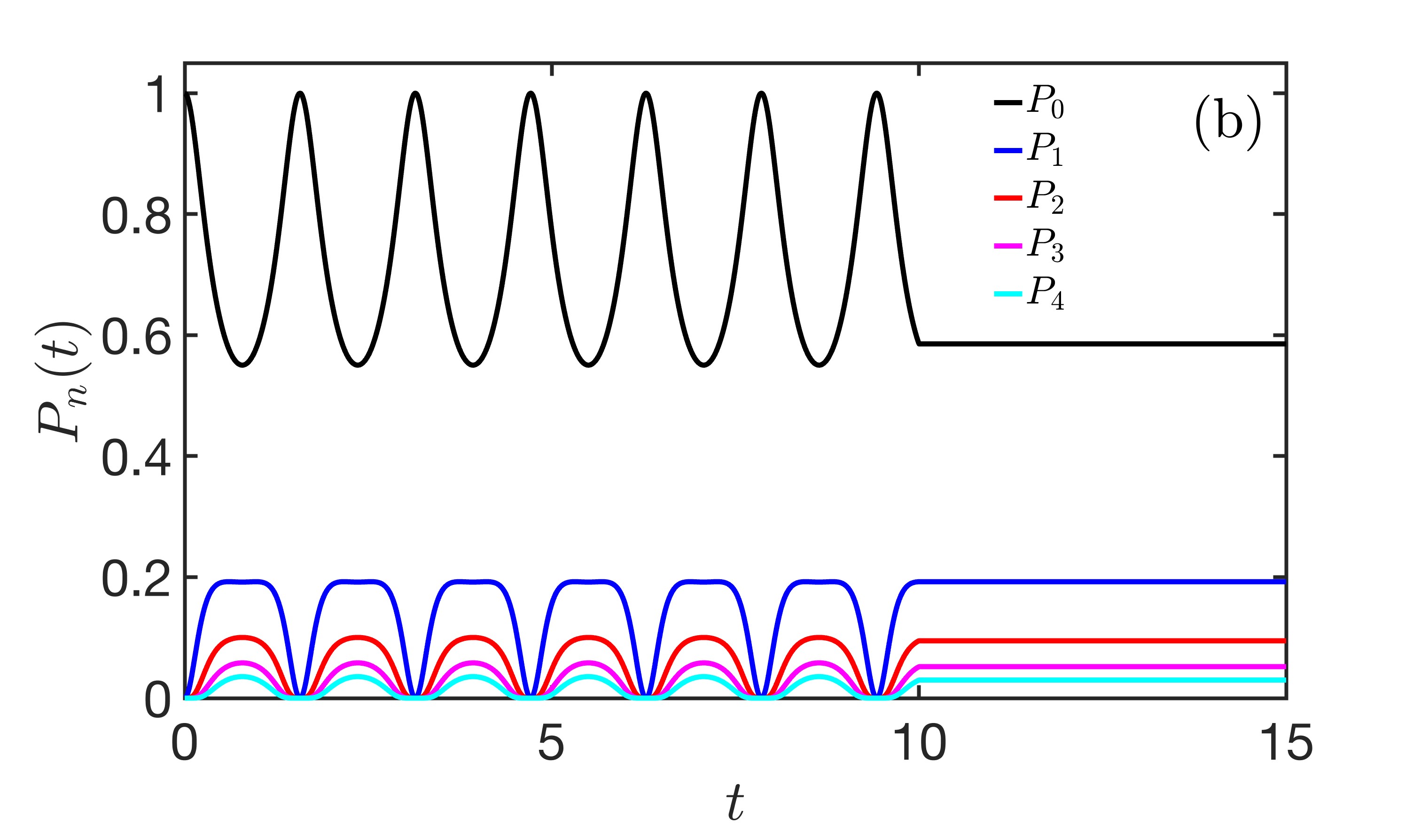}
\caption{(a) Krylov complexity as a function of time for several post-quench frequencies \(\omega_1\) at fixed \(\omega_0=1\). (b) Corresponding Krylov occupation probabilities, displaying the same oscillatory structure ($\omega_0=0.6,\,\omega_1=2,\,\tau=10$).}
\label{fig:HO_dil}
\end{figure}

From the general \(\mathfrak{su}(1,1)\) result with \(\lambda_\alpha=\frac{1}{4}\), the Krylov wavefunction and occupation probabilities are
\begin{equation}
    \varphi_n(t)
    =
    e^{\eta(t)/4}\,z^n(t)
    \sqrt{\frac{(2n)!}{4^n(n!)^2}}\,,
    \qquad
    P_n(t)
    =
    \frac{(2n)!}{4^n(n!)^2}
    \,\bigl(1-|z(t)|^2\bigr)^{1/2}\,|z(t)|^{2n},
\end{equation}
respectively. Therefore, the time-dependent Krylov complexity becomes
\begin{equation}
    K(t)
    =
    \frac{2f_0^2\sin^2\omega_1 t}{\omega_1^2}
    =
    \frac{(\omega_1^2-\omega_0^2)^2}{8\omega_0^2\omega_1^2}
    \sin^2\omega_1 t,
    \qquad 0<t<\tau,
\end{equation}
oscillating at the quenched frequency \(\omega_1\) with an amplitude set by the frequency mismatch. For \(t>\tau\) the coupling vanishes, \(z\) freezes at \(z(\tau)\), and the complexity saturates at \(K(t>\tau)=K(\tau)\).
The real and imaginary parts of the Krylov wavefunction are shown in Fig.~\ref{fig:HO_dil_phi}a and Fig.~\ref{fig:HO_dil_phi}b for a representative quench protocol oscillating during the interval \(0<t<\tau\) and converging to constant values once the quench is switched off. In Fig.~\ref{fig:HO_dil}a and Fig.~\ref{fig:HO_dil}b, we show, respectively, the Krylov complexity for different values of \(\omega_1\) at fixed \(\omega_0\) and the corresponding Krylov occupation probabilities.

\subsection{Virasoro subalgebra}

As a further non-compact example, we consider the rank-one subalgebras obtained by restricting the Virasoro algebra
\begin{equation}
    [L_n,L_m]=(n-m)L_{n+m}+\frac{c}{12}n(n^2-1)\delta_{n+m,0}
\end{equation}
to the modes \(\{L_{-k},L_0,L_k\}\) with fixed \(k>0\), with commutators given by
\begin{equation}
    [L_0,L_{\pm k}]=\mp k L_{\pm k},
    \qquad
    [L_{-k},L_k]=-2kL_0-\frac{c}{12}k(k^2-1).
\end{equation}
Introducing the rescaled generators
\begin{equation}
    J_+=\frac{L_{-k}}{k},
    \qquad
    J_-=\frac{L_k}{k},
    \qquad
    J_0=\frac{L_0}{k}+\frac{c}{24k}(k^2-1),
\end{equation}
one obtains an embedded \(\mathfrak{su}(1,1)\) algebra,
\begin{equation}
    [J_+,J_-]=-2J_0,
    \qquad
    [J_0,J_\pm]=\pm J_\pm.
\end{equation}

A Hamiltonian confined to this sector may be written as
\begin{equation}
    H(t)=kg(t)J_0+kf(t)J_++kf^*(t)J_-.
\end{equation}
Passing to the interaction picture with respect to the Cartan part gives
\begin{equation}
    H_I(t)=\gamma(t)J_+ + \gamma^*(t)J_-,
    \qquad
    \gamma(t)\coloneqq k\,f(t)\exp\!\left(ik\int_0^t\dd t'\,g(t')\right).
\end{equation}

Denoting a Virasoro primary state representing the lowest-weight state by \(\ket{h}\),
\begin{equation}
    J_-\ket{h}=0,
    \qquad
    J_0\ket{h}=\lambda_k\ket{h},
    \qquad
    \lambda_k=\frac{h}{k}+\frac{c}{24k}(k^2-1).
\end{equation}
The associated Krylov states and Lanczos coefficients become
\begin{equation}
    \ket{K_n}
    =
    \frac{J_+^n}{\sqrt{\bra{h}J_-^nJ_+^n\ket{h}}}\ket{h}
    =
    \frac{L_{-k}^n}{\sqrt{\bra{h}L_k^nL_{-k}^n\ket{h}}}\ket{h},\qquad
    b_n=\sqrt{n(2\lambda_k+n-1)},
\end{equation}
by which the Krylov amplitudes satisfy
\begin{equation}
    i\partial_t\varphi_n(t)
    =
    \gamma(t)b_n\,\varphi_{n-1}(t)
    +
    \gamma^*(t)b_{n+1}\,\varphi_{n+1}(t).
\end{equation}

The exact solution follows directly from the general \(\mathfrak{su}(1,1)\) Wei--Norman construction discussed in the main text. Writing
\begin{equation}
    U_I(t)=e^{z(t)J_+}e^{\eta(t)J_0}e^{w(t)J_-},
\end{equation}
with \(z,\eta,w\) determined by the Wei--Norman equations~\eqref{eq:Wei_Norman_eqs} for \(\sigma=-1\), one finds for the Krylov wavefunction
\begin{equation}
    \varphi_n(t)
    =
    e^{\lambda_k\eta(t)}z^n(t)
    \sqrt{\frac{\Gamma(2\lambda_k+n)}{n!\,\Gamma(2\lambda_k)}}.
\end{equation}
Finally, the Krylov occupation probability and complexity become
\begin{equation}
    P_n(t)
    =
    \frac{\Gamma(2\lambda_k+n)}{n!\,\Gamma(2\lambda_k)}
    \bigl(1-|z(t)|^2\bigr)^{2\lambda_k}|z(t)|^{2n},\quad
    K(t)=\frac{2\lambda_k|z(t)|^2}{1-|z(t)|^2},
\end{equation}
respectively.

\subsection{Spin in rotating magnetic field}

As a compact example, we consider a spin in a rotating magnetic field, which realizes the \(\mathfrak{su}(2)\) case in a particularly familiar setting,
\begin{equation}
    H(t)=h(t)\,\mathbf{n}(t)\cdot\boldsymbol{S}
    =
    \frac{h(t)}{2}\sin\theta_2(t)\,e^{-i\theta_1(t)}\,S_+
    +
    \frac{h(t)}{2}\sin\theta_2(t)\,e^{i\theta_1(t)}\,S_-
    +
    h(t)\cos\theta_2(t)\,S_z,
\end{equation}
with \(\mathbf{n}=(\sin\theta_2\cos\theta_1,\,\sin\theta_2\sin\theta_1,\,\cos\theta_2)\).
Comparing with the general formulation~\eqref{eq:H_general}, one identifies
\begin{equation}
    f(t)=\frac{h(t)}{2}\sin\theta_2(t)\,e^{-i\theta_1(t)},
    \qquad
    H_C(t)=h(t)\cos\theta_2(t)\,S_z.
\end{equation}
Passing to the interaction picture with respect to \(H_C(t)\) and using \([S_z,S_+]=S_+\) yields
\begin{equation}
    H_I(t)=\gamma(t)\,S_++\gamma^*(t)\,S_-,
    \qquad
    \gamma(t)=f(t)\,\exp\!\left(i\int_0^t\!\mathrm dt'\,
    h(t')\cos\theta_2(t')\right).
\end{equation}
In the lowest-weight representation with spin \(j\), the time-independent Krylov states are \(\ket{K_n}=\ket{j,-j+n}\), \(n=0,1,\dots,2j\), and the Krylov probabilities and complexity are given by
\begin{equation}
    P_n(t)=\binom{2j}{n}\frac{|z(t)|^{2n}}{(1+|z(t)|^2)^{2j}},
    \qquad
    K(t)=\frac{2j\,|z(t)|^2}{1+|z(t)|^2},
\end{equation}
respectively.

As a solvable example, we set
\begin{equation}
    \theta_2(t)=\theta_0,\qquad \theta_1(t)=\Omega\,t,\qquad h(t)=1,
\end{equation}
so that \(\gamma(t)=\frac{\sin\theta_0}{2}\,e^{-i\delta t}\) with the detuning $\delta\coloneqq\Omega-\cos\theta_0$.

\begin{figure}[t]
\centering
\includegraphics[width=.47\linewidth,trim={2.5cm 0 7cm 0},clip]{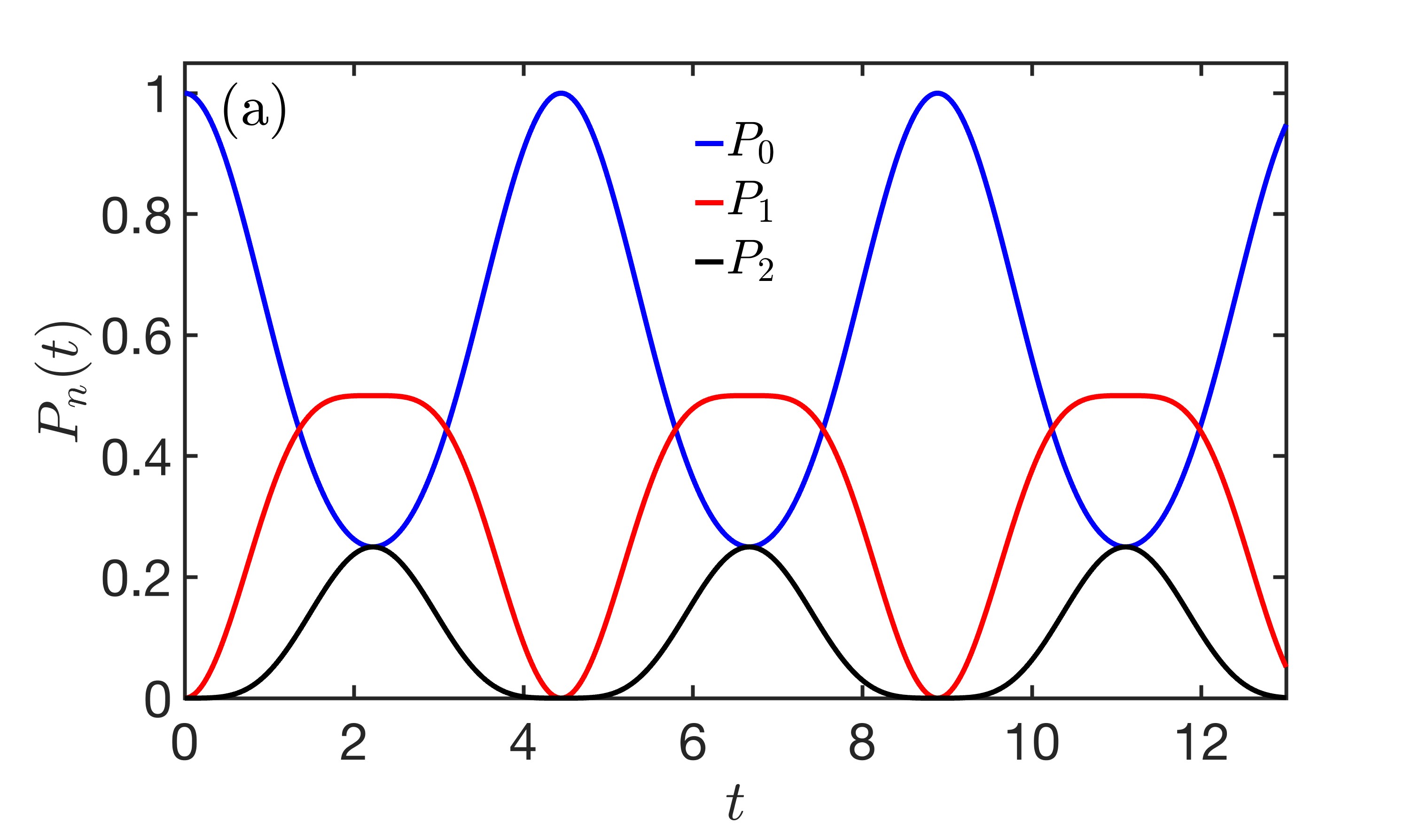}
\includegraphics[width=.47\linewidth,trim={2.5cm 0 7cm 0},clip]{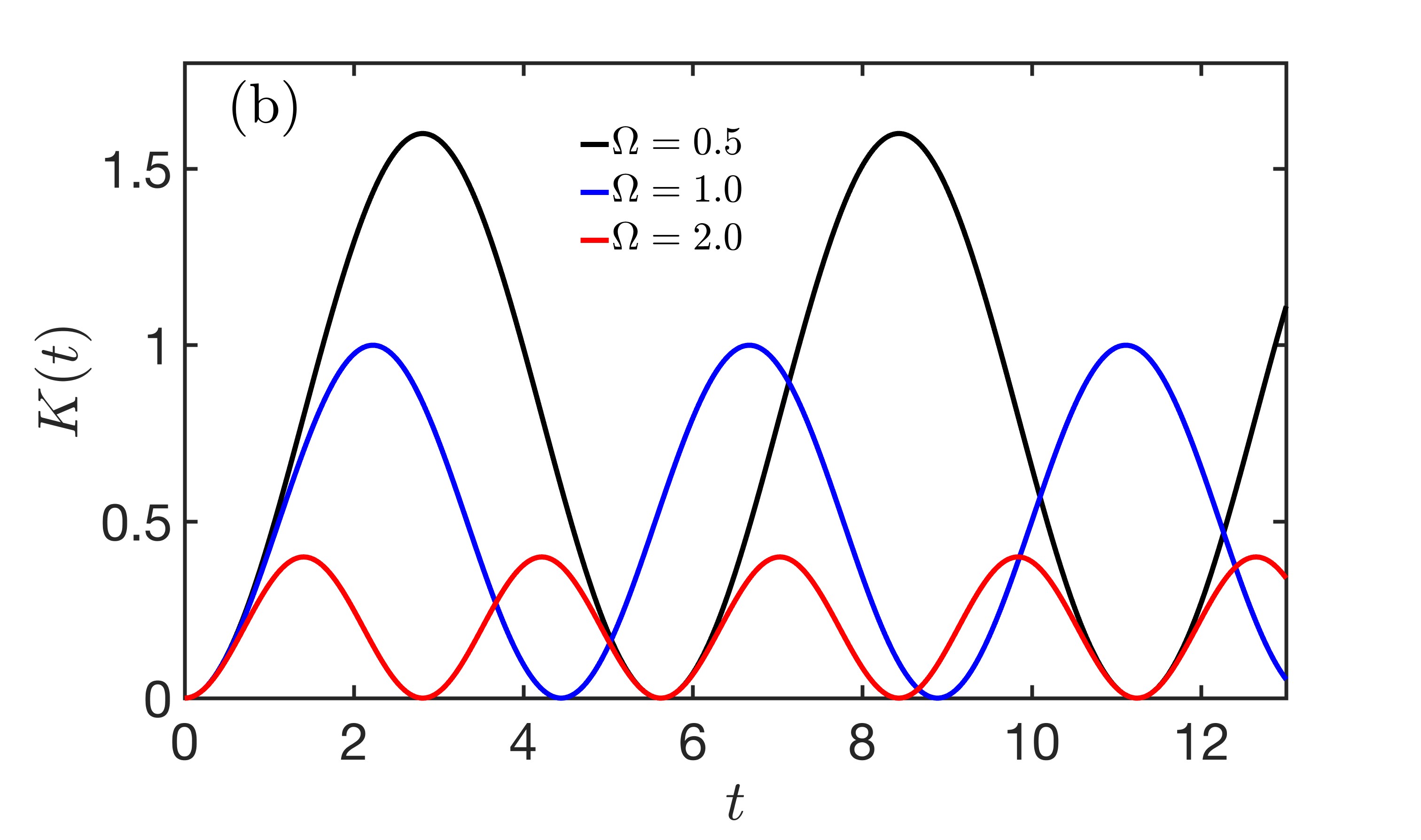}
\caption{(a) Krylov occupation probabilities for the \(\mathfrak{su}(2)\) example, with all occupation probabilities showing oscillatory behavior for \(\Omega=1\). (b) Krylov complexity exhibiting the corresponding oscillatory characteristics in line with the analytical predictions for \(\Omega=0.5,1,2\).}
\label{fig:su_2}
\end{figure}

Then the Wei--Norman equations~\eqref{eq:Wei_Norman_eqs} with \(\sigma=1\) are solved exactly by
\begin{align}
    z(t)&=
    \frac{-i\sin\theta_0\,e^{-i\delta t}\,
    \sin(\Omega_R t/2)}
    {\Omega_R\cos(\Omega_R t/2)
    -i\delta\,\sin(\Omega_R t/2)},
    \\[6pt]
    \eta(t)&=-i\delta\,t
    -2\ln\!\left[
    \cos(\Omega_R t/2)
    -i\frac{\delta}{\Omega_R}\,
    \sin(\Omega_R t/2)
    \right],
    \\[6pt]
    w(t)&=
    \frac{-i\sin\theta_0\,
    \sin(\Omega_R t/2)}
    {\Omega_R\cos(\Omega_R t/2)
    -i\delta\,\sin(\Omega_R t/2)},
\end{align}
where
\begin{equation}
    \Omega_R \coloneqq \sqrt{\delta^2 + \sin^2 \theta_0}
\end{equation}
is the generalized Rabi frequency.
For the numerical demonstration below, we choose the spin-\(1\) representation, so that the Krylov space consists of three states \(n=0,1,2\), and we consider \(\theta_0=\pi/2\), for which \(\delta=\Omega\), \(\Omega_R=\sqrt{\Omega^2+1}\), and the only relevant parameter simplifies to
\begin{equation}
z(t) = \dfrac{-ie^{-i\Omega t}\sin(\sqrt{\Omega^2+1}t/2)}{\sqrt{\Omega^2 + 1}\cos(\sqrt{\Omega^2+1}\,t/2) - i\Omega \sin(\sqrt{\Omega^2+1}\,t/2)}.
\end{equation}
The resulting Krylov occupation probabilities and Krylov complexity are shown in Fig.~\ref{fig:su_2} for different values of \(\Omega\). For \(\Omega=1\), the probabilities \(P_n(t)\) with \(n=0,1,2\) exhibit the expected oscillatory behaviour, while the complexity shows the corresponding periodic response with a frequency that depends on \(\Omega\).

\subsection{Heisenberg--Weyl and oscillator algebra}

The above construction can also be extended beyond simple Lie algebras.
As a representative example, we consider the \(\mathfrak{h}_3\) Heisenberg--Weyl algebra as the simplest example for nilpotent algebras for which the commutator is central and the dynamics reduces to the familiar displacement of the harmonic oscillator. To this end, we consider the general Hamiltonian
\begin{equation}\label{eq:H_H_3}
    H(t)=\gamma^*(t)a+\gamma(t)a^\dagger,
\end{equation}
with \([a,a^\dagger]=1\). This immediately identifies the Krylov states and Lanczos coefficients as
\begin{eqnarray}\label{eq: K_n_HW}
    \ket{K_n}
    =
    \frac{(a^\dagger)^n}{\sqrt{\bra{0}a^n(a^\dagger)^n\ket{0}}}\ket{0}
    =
    \ket{n},
    \qquad
    b_n=\sqrt{n},
\end{eqnarray}
respectively. Accordingly, the Krylov wavefunction satisfies
\begin{eqnarray}
    i\partial_t\varphi_n(t)=\gamma(t)b_n\varphi_{n-1}(t)+\gamma^*(t)b_{n+1}\varphi_{n+1}(t).
\end{eqnarray}
Since
\begin{eqnarray}
    [H(t_1),H(t_2)]
    =
    \gamma^*(t_1)\gamma(t_2)-\gamma(t_1)\gamma^*(t_2)
\end{eqnarray}
is a constant, the time-evolution operator can be written exactly as
\begin{eqnarray}
    U_I(t)
    &=&
    e^{i\Phi(t)}
    e^{-i\alpha(t)a^\dagger-i\alpha^*(t)a}
    =
    e^{i\Phi(t)-|\alpha(t)|^2/2}
    e^{-i\alpha(t)a^\dagger}e^{-i\alpha^*(t)a},
\end{eqnarray}
where
\begin{eqnarray}
    \alpha(t)=\int_0^t\dd t'\,\gamma(t'),
    \qquad
    \Phi(t)=\int_0^t\dd t_1\int_0^{t_1}\dd t_2\,
    \Im\!\bigl[\gamma(t_1)\gamma^*(t_2)\bigr].
\end{eqnarray}
For the vacuum initial state, \(a\ket{0}=0\), this gives
\begin{eqnarray}
    \ket{\psi_I(t)}
    =
    U_I(t)\ket{0}
    =
    e^{i\Phi(t)-|\alpha(t)|^2/2}
    \sum_{n=0}^\infty \frac{(-i)^n\alpha^n(t)}{\sqrt{n!}}\ket{K_n},
\end{eqnarray}
so that
\begin{eqnarray}\label{eq:HW_phi}
    \varphi_n(t)=e^{i\Phi(t)-|\alpha(t)|^2/2}\frac{(-i)^n\alpha^n(t)}{\sqrt{n!}},
    \qquad
    |\varphi_n(t)|^2=e^{-|\alpha(t)|^2}\frac{|\alpha(t)|^{2n}}{n!}.
\end{eqnarray}
Hence, the Krylov complexity is
\begin{eqnarray}
    K(t)=\sum_{n\ge 0}n\,|\varphi_n(t)|^2=|\alpha(t)|^2.
\end{eqnarray}
Equivalently, since the commutator of \(a\) and \(a^\dagger\) is central, the time-evolution operator can be written in a simpler exponential form,
\begin{equation}
    U_I(t)=e^{-iG(t)},
    \qquad
    G(t)=
    -\Phi(t)
    +\alpha(t)\,a^\dagger
    +\alpha^*(t)\,a.
\end{equation}
Its action on the Krylov basis states defined in Eq.~\eqref{eq: K_n_HW} is
\begin{equation}
    G(t)\ket{K_n}
    =
    -\Phi(t)\ket{K_n}
    +\alpha(t)\sqrt{n+1}\ket{K_{n+1}}
    +\alpha^*(t)\sqrt{n}\ket{K_{n-1}},
\end{equation}
so that
\begin{equation}
    a_n(t)=-\Phi(t),
    \qquad
    \tilde b_n(t)=\alpha(t)\sqrt{n}.
\end{equation}
Similar to the $\mathfrak{su}(2)$ and $\mathfrak{su}(1,1)$ simple Lie algebras in Sec.~\ref{sec: Krylov_Wavefunction}, \(G(t)\) is defined only locally at each physical time \(t\), and acts as a time-independent Krylov Liouvillian along a fictitious-time evolution with respect to \(s\),
\[
i\partial_s\psi_n(s)=a_n\,\psi_n+\tilde b_{n+1}\,\psi_{n+1}
+\tilde b_n^*\,\psi_{n-1},\qquad \psi_n(0)=\delta_{n,0}.
\]
Since \(G\) is \(s\)-independent, the solution is \(\ket{\psi(s)}=e^{-isG}\ket{0}\), which by the Baker--Campbell--Hausdorff disentangling gives
\begin{equation}
    \psi_n(s)
    =
    e^{is\Phi-s^2|\alpha|^2/2}\,
    \frac{(-is\alpha)^n}{\sqrt{n!}}\,.
\end{equation}
At \(s=1\), this reproduces the exact physical Krylov wavefunction, \(\psi_n(1)=\varphi_n(t)\), confirming the equivalence of the time-independent generator approach with the full time-dependent dynamics, as in the rank-one case.

If \(\gamma(t)=e^{i\delta}r(t)\) with \(r(t)\in\mathbb R\), then
\begin{equation}
    \Im\!\bigl[\gamma(t_1)\gamma^*(t_2)\bigr]=0,
    \qquad
    \Phi(t)=0,
\end{equation}
and the single-exponential generator becomes purely off-diagonal,
\begin{equation}
    G(t)=\alpha(t)\,a^\dagger+\alpha^*(t)\,a,
    \qquad
    \alpha(t)=e^{i\delta}\int_0^t\dd t'\,r(t').
\end{equation}
Thus, for a fixed phase of \(\gamma(t)\), the generator \(G(t)\) leads to the same Krylov subspace and the same \(\sqrt{n}\) structure as the direct construction, but with off-diagonal coefficients involving the time integral of \(\gamma(t)\) rather than \(\gamma(t)\) itself.

As an example, we consider the translated harmonic oscillator
\begin{equation}
    H(t)=\omega\, a^\dagger a
    -\sqrt{\frac{m\omega^3}{2}}\,x_0(t)\bigl(a+a^\dagger\bigr)
    +E_0(t),
    \qquad
    E_0(t)=\frac{1}{2}m\omega^2x_0^2(t)+\frac{\omega}{2}\,.
\end{equation}
Although this system is described by the oscillator algebra extending \(\mathfrak{h}_3\) with the number operator \(a^\dagger a\), passing to the interaction picture with respect to the latter gets back to \(\mathfrak{h}_3\), with the interaction-picture Hamiltonian
\begin{equation}
    H_I(t)=\gamma^*(t)\,a+\gamma(t)\,a^\dagger+E_0(t),
    \qquad
    \gamma(t)=f(t)\,e^{i\omega t},
    \qquad
    f(t)=-\sqrt{\frac{m\omega^3}{2}}\,x_0(t).
\end{equation}
The scalar term \(E_0(t)\) contributes only an overall phase and therefore does not affect the Krylov probabilities or the complexity. Compared to the general Heisenberg--Weyl Hamiltonian~\eqref{eq:H_H_3}, the Krylov basis remains the Fock basis,
\begin{equation}
    \ket{K_n}=\ket{n},\qquad
    \gamma(t)\sqrt{n}
    =
    -\sqrt{\frac{m\omega^3}{2}}\,x_0(t)\,e^{i\omega t}\sqrt{n}\,,
\end{equation}
and
\begin{equation}
    \alpha(t)=\int_0^t\dd t'\,\gamma(t')
    =
    -\sqrt{\frac{m\omega^3}{2}}
    \int_0^t\dd t'\,x_0(t')\,e^{i\omega t'}.
\end{equation}
From the general Heisenberg--Weyl solution~\eqref{eq:HW_phi}, the Krylov amplitudes and occupation probabilities become
\begin{equation}\label{eq:phi_H_3}
    \varphi_n(t)=e^{i\Phi(t)-|\alpha(t)|^2/2}
    \frac{(-i)^n\alpha^n(t)}{\sqrt{n!}},\quad
    |\varphi_n(t)|^2=e^{-|\alpha(t)|^2}
    \frac{|\alpha(t)|^{2n}}{n!}\,,
\end{equation}
and the Krylov complexity is given by
\begin{equation}
    K(t)=\sum_{n\ge0}n\,|\varphi_n(t)|^2=|\alpha(t)|^2
    =
    \frac{m\omega^3}{2}
    \left|\int_0^t\dd t'\,x_0(t')\,e^{i\omega t'}\right|^2.
\end{equation}

\begin{figure}[t]
\centering
\includegraphics[width=.47\linewidth,trim={1.38cm 0 7cm 0},clip]{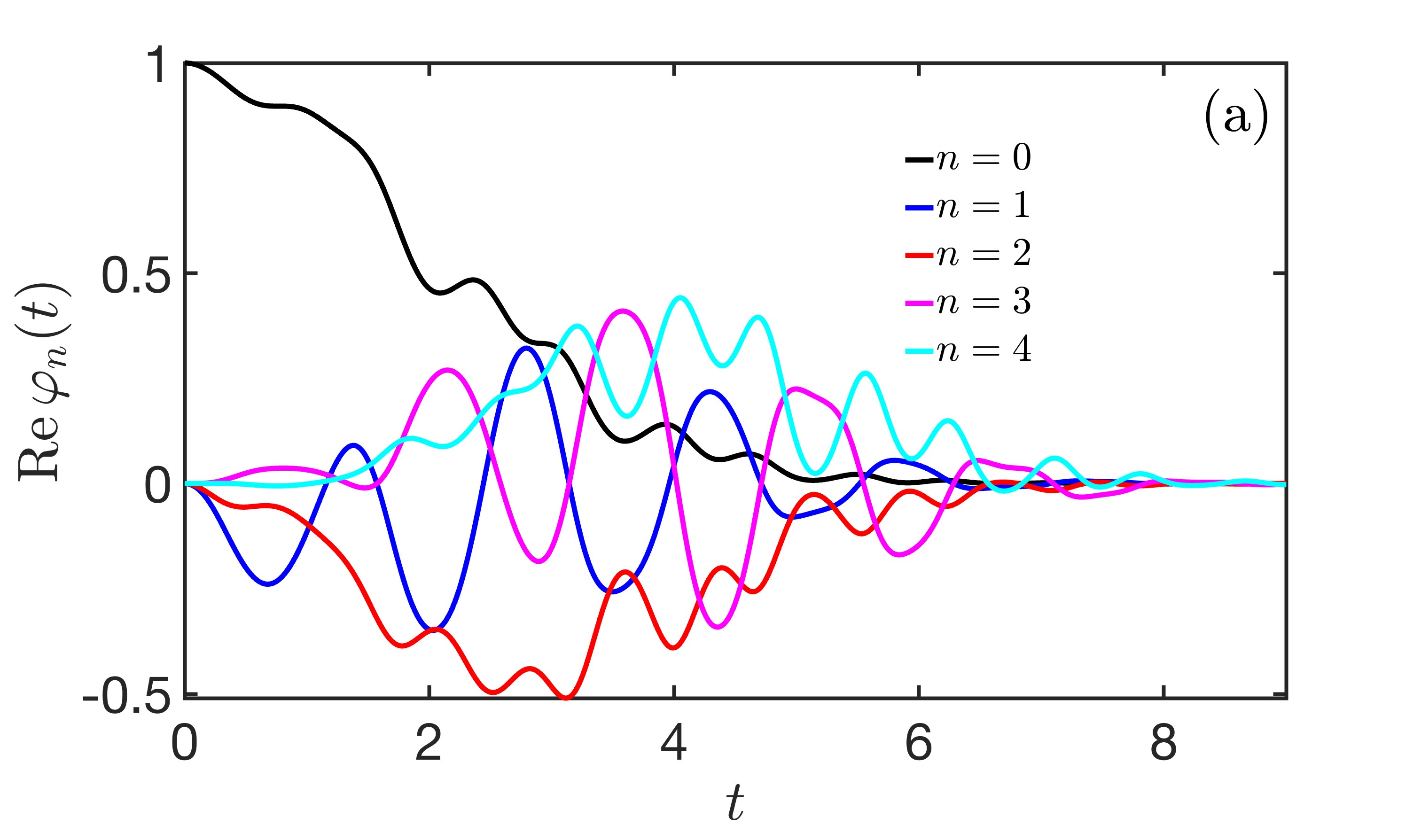}
\includegraphics[width=.47\linewidth,trim={1.38cm 0 7cm 0},clip]{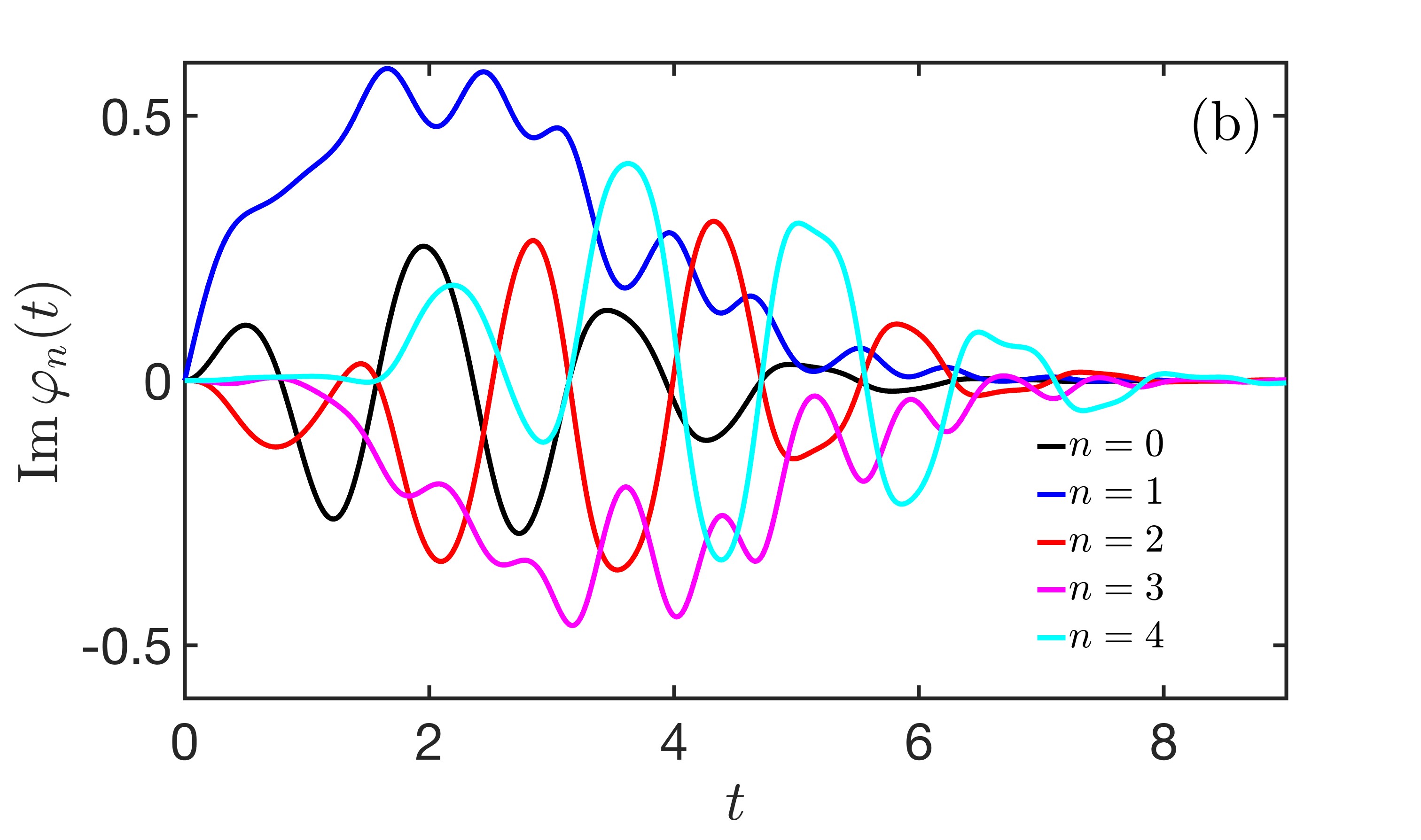}
\caption{(a) Real part of the Krylov wavefunction exhibiting an exponential decay with additional oscillations. (b) Similar features for the imaginary part ($x_0=0.5,\,\omega=2$).}
\label{fig:HO_phi}
\end{figure}

\begin{figure}[t]
\centering
\includegraphics[width=.47\linewidth,trim={1.0cm 0 3cm 0},clip]{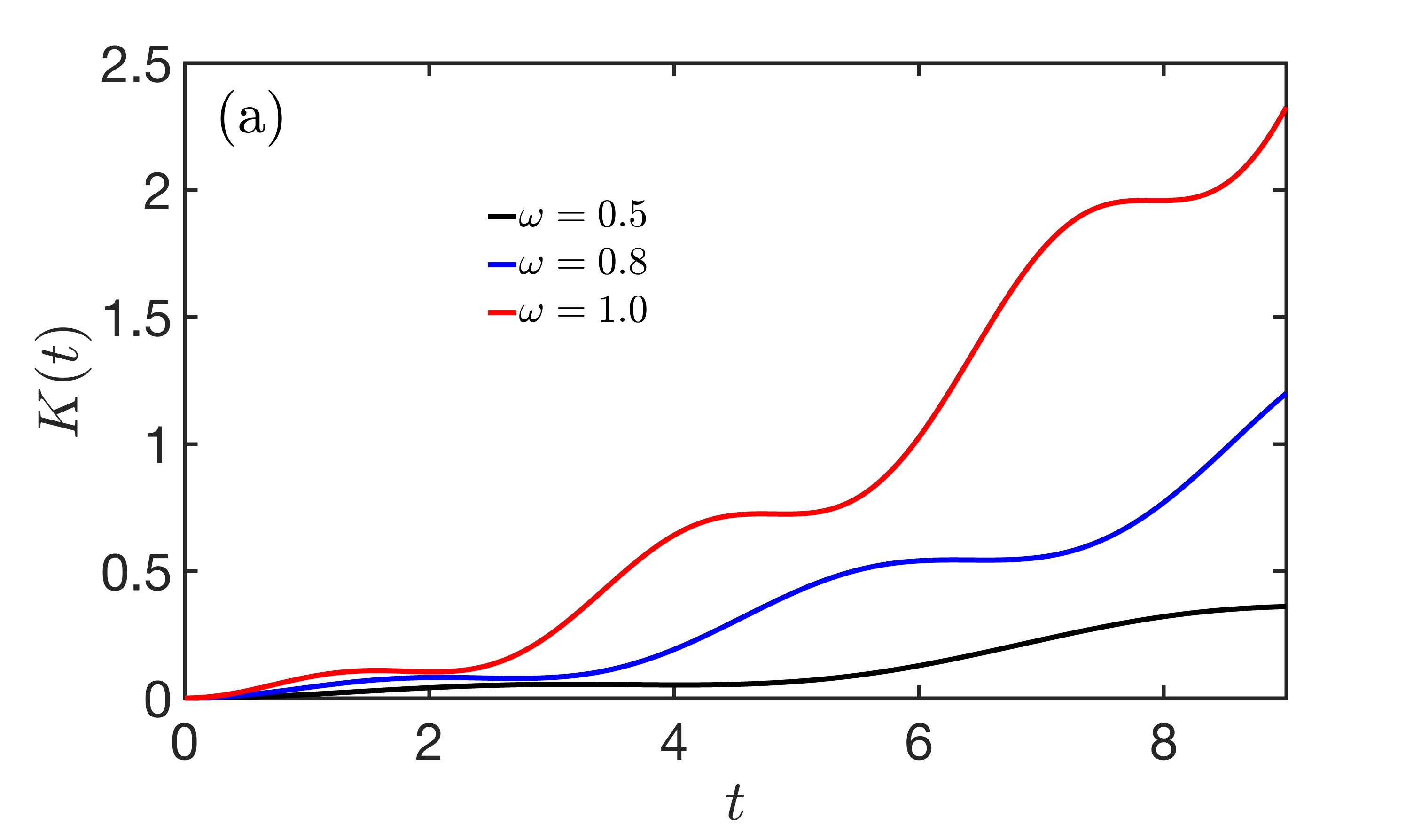}
\includegraphics[width=.47\linewidth,trim={0.5cm 0 3cm 0},clip]{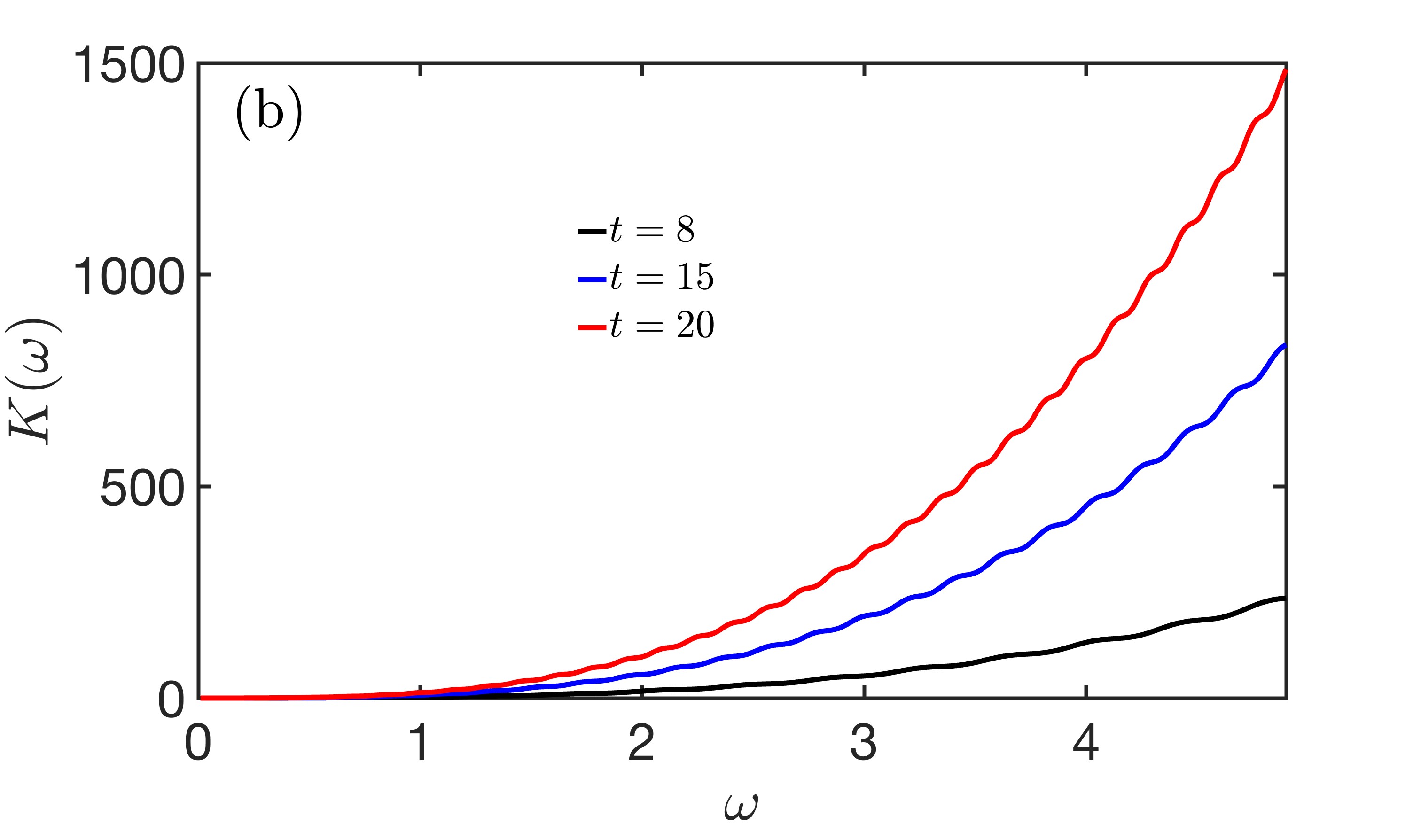}
\caption{(a) Krylov complexity as a function of time for different frequency modulations, showing faster quadratic increase for larger \(\omega\) with additional oscillations. (b) Krylov complexity for fixed instantaneous times with different frequencies, showing cubic growth with the frequency, together with additional oscillations.}
\label{fig:HO}
\end{figure}

To illustrate these results, we consider a harmonically dragged oscillator with \(x_0(t)=x_0\cos(\omega t)\). One finds
\begin{eqnarray}
    \alpha(t)&=&-x_0\sqrt{\frac{m\omega^3}{2}}
    \left[
    \frac{t}{2}
    +\frac{e^{2i\omega t}-1}{4i\omega}
    \right],\\
    K(t)&=&
    \frac{m\omega^3x_0^2}{8}
    \left[
    t^2+\frac{t}{\omega}\sin(2\omega t)
    +\frac{\sin^2(\omega t)}{\omega^2}
    \right].
\end{eqnarray}
The leading \(t^2\) growth reflects the secular energy injection by the resonant component of the drive.
Since \(\gamma(t)=f(t)\,e^{i\omega t}\) carries a time-dependent phase, the equivalent single-exponential generator acquires the additional scalar term \(-\Phi(t)\), although this does not affect the Krylov basis or the complexity.
Numerical results for the real and imaginary parts of the Krylov wavefunction are shown in Fig.~\ref{fig:HO_phi}. The real and imaginary parts exhibit an exponential decay with additional oscillations on top of it, as is obvious from the Poissonian form, Eq.~\eqref{eq:phi_H_3}. The Krylov complexities are plotted in Fig.~\ref{fig:HO}, showing quadratic \(\sim t^2\) and cubic \(\sim \omega^3\) growth for fixed frequencies and instantaneous times, respectively.

\subsection{Higher-dimensional generalization}

The construction developed above extends naturally to the case of several mutually commuting ladder sectors. Let \(\{\alpha_k\}_{k=1}^N\subset \Delta_+\) be a set of positive roots such that, for \(i\neq k\), none of \(\pm\alpha_i\pm\alpha_k\) is a root. Then the corresponding embedded rank-one subalgebras commute, and the interaction-picture reduction factorizes exactly as in the one-dimensional case.

For each selected root \(\alpha_k\), we introduce
\begin{equation}
    L_{+,k}=E_{\alpha_k},
    \qquad
    L_{-,k}=\sigma_k E_{-\alpha_k},
    \qquad
    L_{0,k}=\frac{1}{2}H_{\alpha_k}.
\end{equation}
For selected simple roots, \(H_{\alpha_k}\) is the corresponding simple coroot. Thus, if \(\alpha_k=\alpha_j\), then \(H_{\alpha_k}=H_j\) and \(L_{0,k}=H_j/2\). The Cartan phases are instead determined by the rows \(A_{j\ell}=\alpha_j(H_\ell)\). The corresponding commutation relations read
\begin{equation}
    [L_{+,k},L_{-,k}]=2\sigma_kL_{0,k},
    \qquad
    [L_{0,k},L_{\pm,k}]=\pm L_{\pm,k},
\end{equation}
and all generators belonging to different sectors commute with one another. Here \(\sigma_k=+1\) corresponds to a compact \(\mathfrak{su}(2)\) sector, while \(\sigma_k=-1\) corresponds to a non-compact \(\mathfrak{su}(1,1)\) sector.

Under the same closure conditions as before, the Hamiltonian can be written as
\begin{equation}
    H(t)=
    \sum_{k=1}^N\Bigl(f_k(t)E_{\alpha_k}+\sigma_k f_k^*(t)E_{-\alpha_k}\Bigr)
    +H_E\!\left(\{E_\beta\}_{\beta\notin\{\alpha_k\}_{k=1}^N},t\right)
    +H_C\!\left(\{H_i\},t\right),
\end{equation}
with
\begin{equation}
    H_C\!\left(\{H_i\},t\right)=\sum_i g_i(t)H_i.
\end{equation}
After removing the Cartan sector and the commuting remainder exactly as in the rank-one construction, the final interaction-picture Hamiltonian becomes
\begin{equation}
    H_I(t)=\sum_{k=1}^N\Bigl(\gamma_k(t)L_{+,k}+\gamma_k^*(t)L_{-,k}\Bigr),
    \qquad
    \gamma_k(t)=e^{\,i\sum_{\ell=1}^r\alpha_k(H_\ell)\int_0^t\mathrm dt'\,g_\ell(t')}f_k(t).
\end{equation}

Let \(\ket{\lambda}\) be a common lowest-weight state satisfying
\begin{equation}
    L_{-,k}\ket{\lambda}=0,
    \qquad
    L_{0,k}\ket{\lambda}=\lambda_{\alpha_k}\ket{\lambda},
    \qquad k=1,\dots,N.
\end{equation}
Since the sectors commute, the Krylov basis factorizes as
\begin{equation}
    \ket{K_{n_1,\dots,n_N}}
    =
    \prod_{k=1}^N
    \frac{L_{+,k}^{\,n_k}}
    {\sqrt{\bra{\lambda}L_{-,k}^{\,n_k}L_{+,k}^{\,n_k}\ket{\lambda}}}
    \ket{\lambda}.
\end{equation}
For each direction \(k\), the corresponding Lanczos coefficients are
\begin{equation}
    \bigl(b_{n_k}^{(k)}\bigr)^2
    =
    -\sigma_k\,n_k\bigl(2\lambda_{\alpha_k}+n_k-1\bigr),
\end{equation}
so that
\begin{equation}
    L_{+,k}\ket{K_{n_1,\dots,n_k,\dots,n_N}}
    =
    b_{n_k+1}^{(k)}\ket{K_{n_1,\dots,n_k+1,\dots,n_N}},
\end{equation}
and
\begin{equation}
    L_{-,k}\ket{K_{n_1,\dots,n_k,\dots,n_N}}
    =
    b_{n_k}^{(k)}\ket{K_{n_1,\dots,n_k-1,\dots,n_N}}.
\end{equation}

Expanding the interaction-picture state as
\begin{equation}
    \ket{\psi_I(t)}
    =
    \sum_{n_1,\dots,n_N}\varphi_{n_1,\dots,n_N}(t)\ket{K_{n_1,\dots,n_N}},
\end{equation}
the Schr\"odinger equation reduces to
\begin{equation}
    i\partial_t\varphi_{n_1,\dots,n_N}(t)
    =
    \sum_{k=1}^N
    \left[
    \gamma_k(t)b_{n_k}^{(k)}\,
    \varphi_{n_1,\dots,n_k-1,\dots,n_N}(t)
    +
    \gamma_k^*(t)b_{n_k+1}^{(k)}\,
    \varphi_{n_1,\dots,n_k+1,\dots,n_N}(t)
    \right].
\end{equation}
Because the sectors commute, the full-time evolution operator and hence the Krylov wavefunction also factorize in a product form,
\begin{equation}
    \varphi_{n_1,\dots,n_N}(t)=\prod_{k=1}^N \varphi_{n_k}^{(k)}(t),
\end{equation}
where each \(\varphi_{n_k}^{(k)}(t)\) is determined by the same one-dimensional Wei--Norman analysis as in the rank-one case.

Accordingly, the occupation probabilities factorize as
\begin{equation}
    P_{n_1,\dots,n_N}(t)=\prod_{k=1}^N P_{n_k}^{(k)}(t),\qquad P^{(k)}_{n_k}=|\varphi^{(k)}_{n_k}|^2.
\end{equation}
For a compact sector \(\sigma_k=+1\), writing \(j_k=-\lambda_{\alpha_k}>0\), one obtains
\begin{equation}
    P_{n_k}^{(k)}(t)
    =
    \binom{2j_k}{n_k}
    \frac{|z_k(t)|^{2n_k}}{\bigl(1+|z_k(t)|^2\bigr)^{2j_k}},
    \qquad
    n_k=0,1,\dots,2j_k,
\end{equation}
whereas for a non-compact sector \(\sigma_k=-1\), writing \(\kappa_k=\lambda_{\alpha_k}>0\), one finds
\begin{equation}
    P_{n_k}^{(k)}(t)
    =
    \frac{\Gamma(2\kappa_k+n_k)}{n_k!\,\Gamma(2\kappa_k)}
    \bigl(1-|z_k(t)|^2\bigr)^{2\kappa_k}|z_k(t)|^{2n_k},
    \qquad
    n_k=0,1,2,\dots.
\end{equation}
The total Krylov complexity therefore decomposes additively,
\begin{equation}
    K(t)=\sum_{k=1}^N K_k(t),\quad
    K_k(t)=\frac{-2\sigma_k\lambda_{\alpha_k}|z_k(t)|^2}{1+\sigma_k|z_k(t)|^2}.
\end{equation}

As a minimal illustration of the multidimensional construction, consider the Lie algebra \(\mathfrak{so}(7,\mathbb C)\) of type \(B_3\), whose Dynkin diagram is
\begin{equation*}
\begin{tikzpicture}[scale=1.5, baseline=-0.5ex]
    \foreach \x/\lab in {1/\alpha_1,2/\alpha_2,3/\alpha_3}
        \draw[fill=white] (\x,0) circle (.1) node[below=3pt] {$\lab$};

    \draw (1.1,0) -- (1.9,0);

    \draw (2.1,0.05) -- (2.9,0.05);
    \draw (2.1,-0.05) -- (2.9,-0.05);
\end{tikzpicture}
\end{equation*}
In the present convention \(A_{jk}=\alpha_j(H_k)\), the Cartan action of the simple roots is encoded by the rows
\begin{equation}
\alpha_1(H_k)=(2,-1,0),\qquad
\alpha_2(H_k)=(-1,2,-2),\qquad
\alpha_3(H_k)=(0,-1,2).
\end{equation}
Thus the Cartan matrix is
\begin{equation}
A_{jk}=
\begin{pmatrix}
2 & -1 & 0\\
-1 & 2 & -2\\
0 & -1 & 2
\end{pmatrix},
\qquad
[H_k,E_{\alpha_j}]=A_{jk}E_{\alpha_j}.
\end{equation}
The simple-root pairs close as
\begin{equation}
[E_{\alpha_i},E_{-\alpha_i}]=H_i,
\end{equation}
where \(H_i\equiv H_{\alpha_i}\) are the simple coroots.
Since \(A_{13}=A_{31}=0\), the two end-root sectors \(\alpha_1\) and \(\alpha_3\) commute, providing a simple realization of two independent rank-one Krylov directions~\cite{DiFrancesco1997}.
For these two simple-root sectors, the embedded generators are
\begin{equation}
    L_{+,1}=E_{\alpha_1},
    \qquad
    L_{-,1}=\sigma_1E_{-\alpha_1},
    \qquad
    L_{0,1}=\frac{1}{2}H_1,
\end{equation}
and
\begin{equation}
    L_{+,3}=E_{\alpha_3},
    \qquad
    L_{-,3}=\sigma_3E_{-\alpha_3},
    \qquad
    L_{0,3}=\frac{1}{2}H_3.
\end{equation}

We therefore consider the Hamiltonian
\begin{equation}
    H(t)=f_{\alpha_1}(t)E_{\alpha_1}+\sigma_1 f_{\alpha_1}^*(t)E_{-\alpha_1}
    +f_{\alpha_3}(t)E_{\alpha_3}+\sigma_3 f_{\alpha_3}^*(t)E_{-\alpha_3}
    +\sum_{i=1}^3g_i(t) H_i.
\end{equation}
After removing the Cartan sector, the interaction-picture Hamiltonian becomes
\begin{equation}
    H_I(t)
    =
    \gamma_1(t)L_{+,1}+\gamma_1^*(t)L_{-,1}
    +\gamma_3(t)L_{+,3}+\gamma_3^*(t)L_{-,3},
\end{equation}
where
\begin{equation}
    \gamma_1(t)=e^{\,i\int_0^t\mathrm dt'\,[\,2g_1(t')-g_2(t')\,]}\,f_{\alpha_1}(t),
    \qquad
    \gamma_3(t)=e^{\,i\int_0^t\mathrm dt'\,[\,-g_2(t')+2g_3(t')\,]}\,f_{\alpha_3}(t).
\end{equation}
Because the two sectors commute, the time-evolution operator factorizes,
\begin{equation}
    U_I(t)=U_{I,1}(t)U_{I,3}(t),
\end{equation}
and each factor is governed by the same one-dimensional Wei--Norman problem discussed above.

For a common lowest-weight state \(\ket{\lambda}\) satisfying
\begin{equation}
    L_{-,1}\ket{\lambda}=0,\qquad
    L_{-,3}\ket{\lambda}=0,
\end{equation}
and \(H_i\ket{\lambda}=\lambda_i\ket{\lambda}\), the corresponding lowest weights, defined with respect to \(L_{0,1}=H_1/2\) and \(L_{0,3}=H_3/2\), are
\begin{equation}
    \lambda_{\alpha_1}=\frac{\lambda_1}{2},
    \qquad
    \lambda_{\alpha_3}=\frac{\lambda_3}{2}.
\end{equation}
The Krylov basis therefore factorizes as \(\ket{K_{n,m}}=\ket{K_n^{(1)}}\otimes\ket{K_m^{(3)}}\), so that
\begin{equation}
    \varphi_{n,m}(t)=\varphi_n^{(1)}(t)\varphi_m^{(3)}(t),
    \qquad
    P_{n,m}(t)=P_n^{(1)}(t)P_m^{(3)}(t),
\end{equation}
and the total Krylov complexity is additive,
\begin{equation}
    K(t)=K_1(t)+K_3(t),
    \qquad
    K_j(t)=\frac{-2\sigma_j\lambda_{\alpha_j}|z_j(t)|^2}{1+\sigma_j|z_j(t)|^2},
    \qquad j=1,3.
\end{equation}
Thus, \(\mathfrak{so}(7,\mathbb C)\) provides a minimal explicit application of the higher-dimensional factorized Krylov dynamics described above.

\section{Quantum speed limits to operator growth and saturation in time-dependent Krylov dynamics}

Quantum speed limits refer to various inequalities that set the minimum time for a process to unfold \cite{GongHamazaki2022,Carabba2024}. Several works have explored quantum speed limits for the Krylov complexity growth in the time-independent setting \cite{HornedalEtAl2022,Fan2022,Hornedal2023geometricoperator,Carabba2024,Gill2025,takahashi2024_TDKC,Nandy25}.
In this section, we derive a quantum speed limit (QSL) to the rate of change of Krylov complexity in the time-dependent Lie-algebraic setting. The bound follows from the same Robertson-type argument as in the time-independent case \cite{HornedalEtAl2022}, but its saturation properties are substantially modified by the driving. In particular, while the lowest-weight coherent-state dynamics saturates the bound identically for time-independent generators, time-dependent driving generally impedes this saturation unless a specific phase-locking condition is satisfied.

\subsection{Time-independent case}

We first recall the time-independent bound for the rate of change of the Krylov complexity \cite{HornedalEtAl2022,Carabba2024} with the same notation as in Sec.~\ref {sec: Time_indep_prelim}. 
For
\begin{equation}
    i\partial_t\lvert\mathcal O(t))=\mathcal L\lvert\mathcal O(t)),
    \qquad
    \lvert\mathcal O(t))=\sum_n\varphi_n(t)\lvert\mathcal O_n),
\end{equation}
with
\begin{equation}
    \hat{\mathcal K}=\sum_n n\,\lvert\mathcal O_n)(\mathcal O_n\rvert,
    \qquad
    K(t)=(\mathcal O(t)\rvert\hat{\mathcal K}\lvert\mathcal O(t)),
\end{equation}
the Heisenberg equation and the Robertson inequality give
\begin{equation}
    \left|\partial_t K(t)\right|
    \leq
    2\,\Delta\mathcal L\,\Delta\mathcal K(t).
\end{equation}
Since \(\mathcal L\) is time independent, \(\Delta\mathcal L\) is conserved. For the Krylov initial state \(\lvert\mathcal O_0)\), this gives
\begin{equation}
    \Delta\mathcal L=b_1,
\end{equation}
and therefore
\begin{equation}\label{eq:TI_QSL_Krylov}
    \left|\partial_t K(t)\right|
    \leq
    2b_1\,\Delta\mathcal K(t).
\end{equation}

For the algebraic Liouvillian
\begin{equation}
    \mathcal L=\alpha(L_++L_-),
\end{equation}
acting on the lowest-weight Krylov sector, the coherent-state evolution saturates this bound  \cite{HornedalEtAl2022}. Explicitly,
\begin{align}
\text{ Heisenberg-Weyl}:\qquad
&K(t)=\alpha^2t^2,
&
\Delta\mathcal K(t)=|\alpha t|,\,
&
b_1=\alpha,
\\
\mathfrak{su}(2):\qquad
&K(t)=2j\sin^2(\alpha t),
&
\Delta\mathcal K(t)=\sqrt{\frac{j}{2}}\,|\sin(2\alpha t)|,\,
&
b_1=\alpha\sqrt{2j},
\\
\mathfrak{su}(1,1):\qquad
&K(t)=2\kappa\sinh^2(\alpha t),
&
\Delta\mathcal K(t)=\sqrt{\frac{\kappa}{2}}\,|\sinh(2\alpha t)|,\,
&
b_1=\alpha\sqrt{2\kappa},
\end{align}
with the lowest weights given by $-j$ for $\mathfrak{su}(2)$, $\sigma=1$ and $\kappa$ for $\mathfrak{su}(1,1)$, $\sigma=-1$.
In all three cases,
\begin{equation}
    \left|\partial_t K(t)\right|
    =
    2b_1\,\Delta\mathcal K(t),
\end{equation}
so the QSL is saturated identically, provided the initial Krylov state is a lowest-weight state, or more generally, as formulated in Ref.~\cite{HornedalEtAl2022}, when the Krylov complexity operator can be embedded into a rank-$1$ complexity algebra.

\subsection{Time-dependent case with simple Lie-algebras}
In the time-dependent case, first we consider the same Hamiltonian as in Eq.~\eqref{eq:Ham_Int}, $H_I(t)=\gamma(t)L_+ + \gamma^*(t)L_-$, where $L_\pm$ generate a representation of either $\mathfrak{su}(2)$ or $\mathfrak{su}(1,1),\,$ and the initial Krylov state is the lowest weight state, $\lvert K_0\rangle=\lvert\lambda\rangle$.

The Krylov complexity operator is the shifted Cartan generator satisfying the commutation relations
\begin{equation}
    [\hat{\mathcal K},L_\pm]=\pm L_\pm,
    \qquad
    [H_I(t),\hat{\mathcal K}]=-\gamma(t)L_++\gamma^*(t)L_-.
\end{equation}
The Heisenberg equation in the interaction picture yields
\begin{eqnarray}
    \partial_t K(t)
    =i\bra{\psi_I(t)}[H_I(t),\hat{\mathcal K}]\ket{\psi_I(t)}
    =2\,\Im[\gamma(t)\,C(t)],
    \qquad
    C(t)\coloneqq\bra{\psi_I(t)}L_+\ket{\psi_I(t)},\qquad
\end{eqnarray}
and the Robertson inequality applied to \(H_I(t)\) and \(\hat{\mathcal K}\) gives the time-dependent dispersion bound
\begin{equation}\label{eq:TD_QSL_Krylov}
    \left|\partial_t K(t)\right|
    \leq
    2\,\Delta H_I(t)\,\Delta\mathcal K(t).
\end{equation}
The complexity variance \((\Delta\mathcal K)^2(t)=\langle\hat{\mathcal K}^2\rangle_t - \langle\hat{\mathcal K}\rangle^2_t\) follows directly from the occupation probabilities of Sec.~\ref{sec:TDK_Lie_gen}. Here, $\langle\dots\rangle_t$ denotes the averaging over the time-evolved state, $\ket{\psi_I(t)}$. For \(\mathfrak{su}(2)\) with lowest weight $-j$, the probabilities \(P_n(t)\) correspond to a binomial distribution with parameters \(N=2j\) and \(p=|z|^2/(1+|z|^2)\), giving
\begin{equation}
    (\Delta\mathcal K)^2(t)=Np(1-p)=\frac{2j\,|z(t)|^2}{\bigl(1+|z(t)|^2\bigr)^2}.
\end{equation}
For \(\mathfrak{su}(1,1)\) with lowest weight $\kappa$, the probabilities follow a negative-binomial distribution with \(r=2\kappa\) and \(p=|z|^2\), giving
\begin{equation}
    (\Delta\mathcal K)^2(t)=\frac{rp}{(1-p)^2}=\frac{2\kappa\,|z(t)|^2}{\bigl(1-|z(t)|^2\bigr)^2}.
\end{equation}
Thus, the two cases unify into a single expression
\begin{equation}\label{eq:DK_z}
    (\Delta\mathcal K)^2(t)
    =\frac{-2\sigma\lambda_\alpha\,|z(t)|^2}{\bigl(1+\sigma|z(t)|^2\bigr)^2},
\end{equation}
with \(z(t)\) governed by the Riccati equation \(\dot z=-i\gamma+i\sigma\gamma^*z^2\) for $\sigma=\pm1$. A similar analysis can be performed for the Hamiltonian variance. From \(\langle H_I^2\rangle_t=|\gamma|^2\langle\{L_+,L_-\}\rangle_t+2\Re[\gamma^2\langle L_+^2\rangle_t]\) and \(\langle H_I\rangle_t=2\Re[\gamma(t)C(t)]\), the anti-commutator and the second moment of $L_+$ give
\begin{equation}
    \langle\{L_+,L_-\}\rangle_t
    =-2\sigma\lambda_\alpha
    +\frac{4\lambda_\alpha(2\lambda_\alpha+1)\,|z(t)|^2}{(1+\sigma|z(t)|^2)^2},
    \qquad
    \langle L_+^2\rangle_t
    =\frac{2\lambda_\alpha(2\lambda_\alpha+1)\,(z^*(t))^2}{(1+\sigma|z(t)|^2)^2}.
\end{equation}
Using the identity \(|\gamma(t)|^2|z(t)|^2+\Re[\gamma^2(t)(z^*(t))^2]=2\Re^2[\gamma(t)z^*(t)]\), all contributions collect into
\begin{equation}\label{eq:DHI_z}
    (\Delta H_I)^2(t)
    =
    -2\sigma\lambda_\alpha|\gamma(t)|^2
    +\frac{8\lambda_\alpha\,\Re^2[\gamma(t)z^*(t)]}
          {(1+\sigma|z(t)|^2)^2}.
\end{equation}
Next, we analyze the saturation condition. A time-independent phase of $\gamma(t)=e^{i\delta}|\gamma(t)|$ makes $H_I(t)$ proportional to a fixed algebra element, and hence $[H_I(t),H_I(t')]=0$. When the phase of $\gamma(t)$ varies in time, the Hamiltonians are generically non-commuting, and saturation can occur only under additional instantaneous phase-locking conditions.
For the coherent states considered here, direct evaluation of the two sides of the Schrödinger--Robertson relation gives
\begin{equation}
    4(\Delta H_I)^2(\Delta\mathcal K)^2 = |\partial_t K|^2 + \langle\{\widetilde H_I,\hat{\widetilde{\mathcal K}}\}\rangle^2,
    \qquad
    \hat{\widetilde{\mathcal K}} \coloneqq \hat{\mathcal K} - \langle \hat{\mathcal K}\rangle,\quad \widetilde H_I \coloneqq H_I - \langle H_I\rangle.
\end{equation}
To evaluate the symmetric covariance, one uses \(\langle L_+\hat{\mathcal K}\rangle=\sum_n\varphi_{n+1}^*\varphi_n b_{n+1}n\), yielding from the exact expression for the wave-function, Eq.~\eqref{eq:varphi_n_sol},
\begin{equation}\label{eq:Y}
    \langle L_+\hat{\mathcal K}\rangle
    =\frac{2\lambda_\alpha(2\lambda_\alpha+1)\,|z(t)|^2\,z^*(t)}{(1+\sigma|z(t)|^2)^2}.
\end{equation}
Inserting \eqref{eq:Y} and the coherent-state value of \(C(t)\) into \(\langle\{\widetilde H_I,\hat{\widetilde{\mathcal K}}\}\rangle_t=4\Re[\gamma\langle L_+\hat{\mathcal K}\rangle]+2\Re[\gamma(t)C(t)](1-2K)\), one finds upon simplifying  that
\begin{equation}\label{eq:symcov}
    \langle\{\widetilde H_I,\hat{\widetilde{\mathcal K}}\}\rangle
    =\frac{4\lambda_\alpha(|z(t)|^2-\sigma)\,\Re[\gamma(t)z^*(t)]}{(1+\sigma|z(t)|^2)^2}.
\end{equation}
The Schr\"odinger--Robertson identity therefore gives
\begin{equation}\label{eq:gap}
    4(\Delta H_I)^2(\Delta\mathcal K)^2 - |\partial_t K|^2
    =\frac{16\lambda_\alpha^2(|z(t)|^2-\sigma)^2\,\Re^2[\gamma(t)z^*(t)]}{(1+\sigma|z(t)|^2)^4}\;\geq\;0.
\end{equation}
The condition $\Re[\gamma(t)z^*(t)]=0$ identifies the generic saturation criterion at a given time. At a single time this condition may be satisfied accidentally, even when the Hamiltonians at different times do not commute. Requiring it to hold persistently over a time interval is much more restrictive. Writing
\begin{eqnarray}
    \gamma(t)&=&e^{i\delta(t)}|\gamma(t)|,\qquad
    z(t)=-i e^{i\delta(t)}\zeta(t),
    \qquad
    \zeta(t)\in\mathbb{R},
\end{eqnarray}
and inserting this ansatz into the Riccati equation gives
\begin{eqnarray}
    \dot\delta(t)\,\zeta(t)&=&0,\qquad
    \dot\zeta(t)=|\gamma(t)|\bigl(1+\sigma\zeta^2(t)\bigr).
\end{eqnarray}
Hence, on any interval where the drive is nonzero and $z(t)\neq0$, persistent phase locking forces $\dot\delta(t)=0$. Thus, persistent saturation through this generic evolution requires a time-independent phase of $\gamma(t)$ on that interval. Equivalently,
\begin{eqnarray}
    [H_I(t),H_I(t')]
    =
    2\sigma\!\left[\gamma(t)\gamma^*(t')-\gamma^*(t)\gamma(t')\right]L_0,
\end{eqnarray}
so a constant phase of $\gamma(t)$ makes the Hamiltonian commute with itself at different times. Conversely, when the phase of $\gamma(t)$ varies in time, the Hamiltonians are generically non-commuting and the phase-locking condition can survive only at isolated accidental times, not as a persistent saturation mechanism.
In addition, the compact $\mathfrak{su}(2)$ case admits an exceptional instantaneous saturation at all times. From Eq.~\eqref{eq:gap}, for $\sigma=+1$ the right-hand side also vanishes when $|z(t)|=1$, independently of the value of $\Re[\gamma(t)z^*(t)]$. This is a special compact-sector effect and should be understood as an accidental saturation at isolated times rather than as a persistent saturation mechanism.

In the commuting case, the dispersion entering the bound reduces to the instantaneous hopping scale. Indeed, for the lowest-weight initial state one has
\[
(\Delta H_I)^2(t)=\langle\left(H_I-\langle H_I\rangle_t\right)^2\rangle_t=\langle\left(H_I-\langle H_I\rangle_0\right)^2\rangle_t=|\gamma(t)|^2 b_1^2,
\]
so that the time-dependent bound is saturated as
\[
|\partial_t K(t)|= 2|\gamma(t)|b_1\,\Delta\mathcal K(t),
\]
which is the natural time-dependent analogue of the time-independent \(b_1\) bound.

\subsection{Heisenberg--Weyl algebra}
Finally, we consider the Heisenberg--Weyl algebra $\mathfrak{h}_3$, with Hamiltonian
\begin{equation}
    H_I(t)=\gamma^*(t)a+\gamma(t)a^\dagger,
    \qquad [a,a^\dagger]=1.
\end{equation}
For the vacuum initial state, $a\ket{0}=0$, the Krylov wave-function is a coherent state
\begin{eqnarray}
    \varphi_n(t)=
    e^{i\Phi(t)-|\alpha(t)|^2/2}\frac{(-i)^n\alpha^n(t)}{\sqrt{n!}},
\end{eqnarray}
leading to a Poisson distribution over the Krylov basis with the Krylov complexity variance equaling the complexity itself
\begin{eqnarray}
    K(t)=\langle\hat{\mathcal K}\rangle_t=
    (\Delta\mathcal K)^2(t)&=&|\alpha(t)|^2.
\end{eqnarray}
The variance of the Hamiltonian can be obtained from the coherent states as
\begin{eqnarray}
(\Delta H_I)^2
&=&
|\gamma|^2\langle\{a,a^\dagger\}\rangle
+\gamma^2\langle a^{\dagger 2}\rangle
+(\gamma^*)^2\langle a^2\rangle
-\bigl(\gamma^*\langle a\rangle+\gamma\langle a^\dagger\rangle\bigr)^2,
\\
&=&
|\gamma|^2\bigl(\langle\{a,a^\dagger\}\rangle-2|\langle a\rangle|^2\bigr)
=
|\gamma|^2\langle[a,a^\dagger]\rangle
=
|\gamma|^2.
\end{eqnarray}
On the other hand, the growth rate of the Krylov complexity is given by
\begin{eqnarray}
\partial_t K(t)
&=&
\partial_t |\alpha(t)|^2
=
\dot\alpha(t)\alpha^*(t)+\alpha(t)\dot\alpha^*(t),
\\
&=&
\gamma(t)\alpha^*(t)+\gamma^*(t)\alpha(t)
=
2\,\Re\!\left[\gamma(t)\alpha^*(t)\right],
\\
|\partial_t K(t)|^2
&=&
4\,\Re^2\!\left[\gamma(t)\alpha^*(t)\right].
\end{eqnarray}
Thus, the bound reads
\begin{equation}
    |\partial_t K(t)|^2
    \le
    4\,(\Delta H_I)^2(t)\,(\Delta\mathcal K)^2(t)
    =
    4\,|\gamma(t)|^2\,|\alpha(t)|^2.
\end{equation}
As a result, the bound is saturated at a given time when
\(\gamma(t)\alpha^*(t)\in\mathbb{R}\), i.e., when the instantaneous
drive is phase-aligned with the accumulated displacement. Saturation for
all times, starting from \(\alpha(0)=0\), requires this phase alignment
to persist throughout the evolution. This is obtained when
\(\gamma(t)=e^{i\delta}r(t)\), with constant \(\delta\) and real
\(r(t)\), in which case all Hamiltonians are proportional to the same
fixed quadrature and therefore commute at different times,
\([H_I(t),H_I(t')]=0\). For a genuinely time-dependent phase of
\(\gamma(t)\), this persistent alignment is generically lost, although
isolated accidental saturation times are not excluded.

Similarly to the \(\mathfrak{su}(2)\) and \(\mathfrak{su}(1,1)\) cases,
in this commuting case the dispersion reduces to the instantaneous
hopping scale,
\begin{eqnarray}
    \Delta H_I(t)=|\gamma(t)|,
\end{eqnarray}
and the bound is saturated in the instantaneous form
\begin{eqnarray}
    |\partial_t K(t)|
    =
    2|\gamma(t)|\,\Delta\mathcal K(t),
\end{eqnarray}
with \(b_1=1\) for the Heisenberg--Weyl algebra.

 \section{Conclusion}
In this work, we developed a Lie-algebraic framework for describing the quantum dynamics induced by a time-dependent generator in a time-dependent Krylov subspace. Our main result is that, for a broad class of driven Hamiltonians with an underlying algebraic structure, the exact time-dependent dynamics formulated in Krylov space can be organized directly in terms of the ladder operators and the root structure of the corresponding Lie algebra. In particular, we identified the minimal conditions under which the interaction-picture Hamiltonian reduces to an embedded rank-one algebra, yielding an exact one-dimensional Krylov chain governed by an effective \(\mathfrak{su}(2)\) or \(\mathfrak{su}(1,1)\) structure. This yields closed expressions for the Krylov basis, the Lanczos coefficients, the Krylov wavefunction, and the Krylov complexity. We further showed that the Lanczos algorithm applied to the exact exponential representation of the time-evolution operator defines a distinct, effective time-independent Krylov dynamics. In this formulation, the temporal non-locality of the exact generator is no longer manifest at the level of the Krylov recursion, since it is encoded implicitly in the effective exponentiated operator. The resulting evolution in the same Krylov subspace therefore proceeds in a fictitious time, while the physical time enters only parametrically through the effective generator. Nevertheless, the exact time-dependent Krylov wavefunction is recovered by evaluating this auxiliary evolution at unit fictitious time.
The construction also extends naturally to higher-rank algebras with several commuting ladder pairs, where the dynamics acquires the structure of a higher-dimensional Krylov lattice. In that setting, the evolution factorizes into independent algebraic sectors, and the Krylov wavefunction and complexity decompose accordingly. This provides a geometric picture of how time-dependent operator growth is constrained by the root-space structure of the underlying algebra.

Beyond simple Lie algebras, we show that the same framework also applies to nilpotent settings, in particular to the Heisenberg--Weyl algebra and its oscillator extension, as realized for instance in the translated harmonic oscillator.
We illustrate the formalism in several representative physical settings, including the translated and dilated harmonic oscillator, a spin in a rotating magnetic field, closed Virasoro subalgebras, and higher-dimensional multilevel examples. These examples show that the formalism is not only algebraically natural, but also directly applicable to physically relevant driven systems. In each case, the Lie-algebraic structure determines the corresponding Krylov hierarchy and enables an exact characterization of the associated complexity growth.

Finally, we generalized the quantum speed limit to Krylov complexity to time-dependent generators and analyzed it in the algebraic setting for the $\mathfrak{su}(2)$, $\mathfrak{su}(1,1)$, and Heisenberg--Weyl cases. We found that the bound retains the same Robertson form as in the time-independent case, while its saturation is strongly constrained by the driving. Persistent saturation occurs only under the corresponding phase-locking condition, which in the commuting case is realized when the Hamiltonian is proportional to a fixed algebra element at all times.

Overall, our results provide a unified algebraic perspective on dynamics in time-dependent Krylov subspaces and establish a direct link between Krylov complexity and the symmetry structure of the Hamiltonian. They further show that exact nonequilibrium Krylov dynamics is accessible in broad classes of systems whenever the interaction-picture evolution can be organized into closed ladder structures. A natural direction for future work is to relax the present commutativity and closure conditions in order to describe more general interacting many-body systems, non-factorizable higher-rank dynamics, and genuinely non-integrable regimes. It would also be interesting to combine the present framework with approximate algebraic truncation schemes, with the aim of capturing complexity growth in systems where exact Lie closure is only weakly broken, and to extend the construction to more general higher-rank simple Lie algebras associated with non-orthogonal higher-dimensional Krylov lattices.
\\

{\bf Acknowledgements}
This project was supported by the Luxembourg National Research Fund via the grants C22/MS/17132054/AQCQNET and CC25/MS/19559370/FastQOPT.  E.M.-G. was supported by the Deutsche Forschungsgemeinschaft (DFG, German Research Foundation) grant SH 81/8-1 and by a National Science Foundation (NSF)–Binational Science Foundation (BSF) grant 2023666.
\appendix

\section{Time-evolution operator approach to Krylov subspace}\label{app:U_G}

In this appendix, we derive the exact relation between the exponential-generator representation of the interaction-picture time-evolution operator, \(U_I(t)=e^{-iG(t)}\), and its Wei--Norman factorization. We first consider the \(\mathfrak{su}(2)\) and \(\mathfrak{su}(1,1)\) cases, with the general Hamiltonian of Eq.~\eqref{eq:Ham_Int},
\begin{equation}
    H_I(t)=\gamma(t)L_+ + \gamma^*(t)L_-,
    \qquad
    [L_+,L_-]=2\sigma L_0,\qquad [L_0,L_\pm]=\pm L_\pm.
\end{equation}
The lowest-weight state is denoted by
\begin{equation}
    L_-\ket{\lambda}=0,
    \qquad
    L_0\ket{\lambda}=\lambda\ket{\lambda},
\end{equation}
where, for brevity, we suppress the subscript on \(\lambda_\alpha\) and write \(\lambda\equiv\lambda_\alpha\). The corresponding Krylov basis, Lanczos coefficients, Krylov wavefunction, and Krylov complexity are given by (see Sec.~\ref{sec: Krylov_Wavefunction})
\begin{eqnarray}\label{eq:phi_n_app}
    \ket{K_n}&=&\frac{L_+^n}{\sqrt{\bra{\lambda}L_-^nL_+^n\ket{\lambda}}}\ket{\lambda},
    \qquad
    b_n=\sqrt{-\sigma\,n(2\lambda+n-1)},\\
    i\partial_t\varphi_n(t)&=&\gamma(t)b_n\,\varphi_{n-1}(t)+\gamma^*(t)b_{n+1}\,\varphi_{n+1}(t),
    \qquad
    K(t)=\sum_n n\,|\varphi_n(t)|^2.
\end{eqnarray}

The interaction-picture propagator satisfies
\begin{equation}
    i\partial_t U_I(t)=H_I(t)U_I(t),
    \qquad
    U_I(0)=\mathbb{I}.
\end{equation}
As in the main text, the exact logarithm of $U_I(t)$ can be chosen, on each continuous branch, inside the same rank-one algebra,
\begin{equation}\label{eq:UI_single_exp}
    U_I(t)=e^{-iG(t)},
    \qquad
    G(t)=\Theta_0(t)\,L_0+\Theta_+(t)\,L_+ + \Theta_+^*(t)\,L_-,
    \qquad
    \Theta_0\in\mathbb{R}.
\end{equation}
We now derive the relation between \((\Theta_0,\Theta_+)\) and the Wei--Norman parameters \((z,\eta,w)\) defined by
\begin{equation}
    U_I(t)=e^{zL_+}e^{\eta L_0}e^{wL_-}.
\end{equation}

To this end, introduce an auxiliary parameter \(s\in[0,1]\) and define \(U(s)=e^{-isG}\). Differentiating the Wei--Norman factorization and using the adjoint identities
\begin{eqnarray}
    e^{zL_+}L_0\,e^{-zL_+}&=&L_0-zL_+,\\
    e^{zL_+}L_-\,e^{-zL_+}&=&L_-+2\sigma zL_0-\sigma z^2L_+,\\
    e^{\eta L_0}L_-\,e^{-\eta L_0}&=&e^{-\eta}L_-,
\end{eqnarray}
and matching the coefficients in \(\partial_s U\,U^{-1}=-iG\), one obtains~\cite{WeiNorman1963,Charzynski_2013}
\begin{equation}\label{eq:riccati}
    \partial_s z = -i\bigl(\Theta_+ +\Theta_0\,z -\sigma\Theta_+^*\,z^2\bigr),
    \qquad
    \partial_s\eta = -i\Theta_0 +2i\sigma\Theta_+^*\,z,
\end{equation}
with initial conditions \(z(0)=\eta(0)=0\).

For \(\Theta_+\neq 0\), the first equation is a constant-coefficient Riccati equation. The substitution
\begin{equation}
    z=-(\partial_s u)/(i\sigma\Theta_+^*u)
\end{equation}
linearizes it to
\begin{equation}
    \partial_s^2 u+i\Theta_0\,\partial_s u+\sigma|\Theta_+|^2 u=0,
\end{equation}
with \(u(0)=1\) and \(\partial_s u(0)=0\). The characteristic roots to this second-order ordinary differential equation with constant coefficients are given by
\begin{eqnarray}
    \mu_\pm&=&i(-\Theta_0/2\pm\chi),\\
    \chi&=&\sqrt{\sigma|\Theta_+|^2+\tfrac{1}{4}\Theta_0^2}\,,
\end{eqnarray}
and therefore
\begin{equation}
    u(s)=e^{-i\Theta_0 s/2}\left[\cos(\chi s)
    +i\frac{\Theta_0}{2\chi}\sin(\chi s)\right].
\end{equation}
Substituting into the equation for \(\eta\) gives
\begin{equation}
    \partial_s\eta=-i\Theta_0-2\,\frac{\partial_s u}{u},
\end{equation}
which integrates to
\begin{equation}
    e^{-\eta/2}=e^{i\Theta_0 s/2}\,u(s).
\end{equation}
Evaluating at \(s=1\), one finds
\begin{align}\label{eq:disentangle}
    e^{-\eta/2}
    &=\cos\chi+i\frac{\Theta_0}{2\chi}\sin\chi,
    &
    z\,e^{-\eta/2}
    &=-i\frac{\Theta_+}{\chi}\sin\chi,
    &
    w\,e^{-\eta/2}
    &=-i\frac{\Theta_+^*}{\chi}\sin\chi.
\end{align}
For \(\Theta_+=0\), one simply has \(z=w=0\) and \(\eta=-i\Theta_0 s\). In the compact case \((\sigma=1)\), \(\chi\) is real, while in the non-compact case \((\sigma=-1)\), \(\chi\) may become imaginary when \(|\Theta_+|>|\Theta_0|/2\), in which case the above expressions continue analytically to hyperbolic functions through \(\cos(i\kappa)=\cosh\kappa\) and \(\sin(i\kappa)=i\sinh\kappa\).

These relations can be inverted by defining
\begin{equation}
    A\coloneqq e^{-\eta/2},
    \qquad
    B\coloneqq z\,e^{-\eta/2},
\end{equation}
so that \(\operatorname{Re}A=\cos\chi\) and
\begin{equation}
    (\operatorname{Im}A)^2+\sigma|B|^2=\sin^2\chi,
\end{equation}
equivalently,
\begin{equation}
    |A|^2+\sigma|B|^2=1.
\end{equation}
Hence
\begin{align}\label{eq:inversion}
    \chi
    &=\arctan\frac{\sqrt{(\operatorname{Im}A)^2+\sigma|B|^2}}
                   {\operatorname{Re}A},
    &
    \Theta_0
    &=\frac{2\chi\,\operatorname{Im}A}
           {\sqrt{(\operatorname{Im}A)^2+\sigma|B|^2}},
    &
    \Theta_+
    &=\frac{i\chi\,B}
           {\sqrt{(\operatorname{Im}A)^2+\sigma|B|^2}}.
\end{align}
The arctangent is understood as the branch-continuous two-argument arctangent, with the branch fixed by continuity from $\chi(0)=0$. In the non-compact case, $\sigma=-1$, when the argument under the square root becomes negative, the same expression is understood by analytic continuation to imaginary $\chi$. The identity \( |A|^2+\sigma|B|^2=1 \) encodes the unitarity condition for \(\mathrm{SU}(2)\) when \(\sigma=1\) and for \(\mathrm{SU}(1,1)\) when \(\sigma=-1\). Thus, the exact logarithm of \(U_I(t)\) remains in the rank-one algebra but generically acquires a non-vanishing Cartan component \(\Theta_0L_0\) whenever \(\operatorname{Im}[e^{-\eta/2}]\neq0\).

To construct the Krylov basis associated with the exponential representation~$\eqref{eq:UI_single_exp}$, assuming that $\Theta_+(t)\neq0$, one starts the Lanczos algorithm with $G(t)$ from the lowest-weight state $\ket{\lambda}$. This generates the same ladder Krylov states as in Eq.~$\eqref{eq:phi_n_app}$,
\begin{equation}
    G(t)\ket{K_n}
    =
    a_n(t)\ket{K_n}
    +
    \tilde b_{n+1}(t)\ket{K_{n+1}}
    +
    \tilde b_n^*(t)\ket{K_{n-1}},
\end{equation}
where
\begin{equation}
    a_n(t)=\Theta_0(t)(\lambda+n),\qquad
    \tilde b_{n+1}(t)=\Theta_+(t)b_{n+1}
    =\Theta_+(t)\sqrt{-\sigma (n+1)(2\lambda+n)}.
\end{equation}
If $\Theta_+(t)=0$, then $G(t)=\Theta_0(t)L_0$ is diagonal in the Krylov basis and the auxiliary Krylov problem is trivial.
Thus, the generator \(G(t)\) is represented by a tridiagonal Hermitian matrix in the same Krylov basis. The off-diagonal coefficients carry the common phase of \(\Theta_+(t)\), which may be removed by the rephasing \(\ket{K_n}\to e^{in\arg\Theta_+}\ket{K_n}\), bringing the tridiagonal form to the standard real-phase Lanczos gauge whenever convenient.
In \(G(t)\) the physical time \(t\) appears locally and as a fixed parameter, and the evolution is generated with respect to a fictitious time parameter \(s\) according to an effective time-independent Lanczos algorithm,
\begin{equation}
    i\partial_s\ket{\psi_I(s)}=G(t)\ket{\psi_I(s)},
    \qquad
    \ket{\psi_I(0)}=\ket{\lambda}\equiv\ket{0},
    \qquad
    s\in[0,1].
\end{equation}
Expanding
\begin{equation}
    \ket{\psi_I(s)}=\sum_n\psi_n(s)\ket{K_n},
\end{equation}
and using the tridiagonal action of \(G\), one obtains
\begin{equation}\label{eq:s_chain}
    i\partial_s\psi_n(s)
    =
    a_n\,\psi_n(s)
    +
    \tilde b_{n+1}\,\psi_{n+1}(s)
    +
    \tilde b_n^*\,\psi_{n-1}(s),
    \qquad
    \psi_n(0)=\delta_{n,0},
\end{equation}
where \(a_n=\Theta_0(\lambda+n)\) and \(\tilde b_n=\Theta_+b_n\) are constant in \(s\). Since \(G(t)\) is \(s\)-independent, the solution is formally
\begin{equation}
    \ket{\psi_I(s)}=e^{-isG(t)}\ket{0}.
\end{equation}
The disentangling formula~\eqref{eq:disentangle} then applies at any \(s\),
\begin{equation}
    e^{-isG(t)}=e^{Z(s)L_+}e^{\mathcal E(s)L_0}e^{W(s)L_-}.
\end{equation}
Acting on \(\ket{0}\), the rightmost factor drops out because \(L_-\ket{0}=0\), while the middle factor contributes \(e^{\lambda\mathcal E(s)}\). Hence
\begin{equation}
    \psi_n(s)
    =\bra{K_n}\ket{\psi_I(s)}
    =e^{\lambda\mathcal E(s)}\,\frac{Z(s)^n}{n!}
    \sqrt{\bra{0}L_-^nL_+^n\ket{0}}\,.
\end{equation}

To identify the solution at \(s=1\) with the physical Krylov wavefunction, recall that by definition
\begin{equation}
    U_I(t)=e^{-iG(t)},
\end{equation}
while the main text factorizes the same operator as
\begin{equation}
    U_I(t)=e^{z(t)L_+}e^{\eta(t)L_0}e^{w(t)L_-}.
\end{equation}
Setting \(s=1\) in the auxiliary disentangling formula gives
\begin{equation}
    e^{-iG(t)}=e^{Z(1)L_+}e^{\mathcal E(1)L_0}e^{W(1)L_-}.
\end{equation}
Since both factorizations represent the same operator with the unique Wei--Norman decomposition, the corresponding coefficients can be identified,
\begin{equation}
    Z(1)=z(t),
    \qquad
    \mathcal E(1)=\eta(t),
    \qquad
    W(1)=w(t).
\end{equation}
Substituting this into the expression for \(\psi_n(s)\) at \(s=1\) yields
\begin{equation}
    \psi_n(1)
    =e^{\lambda\eta(t)}\,\frac{z(t)^n}{n!}
    \sqrt{\bra{0}L_-^nL_+^n\ket{0}}\,,
\end{equation}
which is precisely \(\varphi_n(t)=\braket{K_n}{U_I(t)\ket{0}}\) obtained from the Wei--Norman solution in the main text. Thus, the auxiliary tridiagonal recursion~\eqref{eq:s_chain}, whose coefficients are determined by the Lanczos algorithm applied to \(G(t)\), reproduces the exact interaction-picture Krylov wavefunction at \(s=1\).

As a consistency check, consider the constant-phase case
\begin{equation}
    \gamma(t)=e^{i\delta}r(t),
\end{equation}
with \(r(t)\) real. Then
\begin{equation}
    H_I(t)=r(t)\bigl(e^{i\delta}L_++e^{-i\delta}L_-\bigr)
\end{equation}
is proportional at all times to a fixed algebra element, so time ordering is trivial, and one immediately reads off
\begin{equation}
    \Theta_0=0,
    \qquad
    \Theta_+=\int_0^t\gamma(s)\,\mathrm ds.
\end{equation}
The same result follows from the inversion formulas~\eqref{eq:inversion}. Indeed, the Wei--Norman equations
\begin{equation}
    \partial_t z=-i\gamma+i\sigma\gamma^*z^2,
    \qquad
    \partial_t\eta=2i\sigma\gamma^*z,
\end{equation}
are solved by the ansatz \(z=e^{i\delta}\zeta\) with \(\zeta\) purely imaginary, which reduces the Riccati equation to
\begin{equation}
    \partial_t\zeta=ir(\sigma\zeta^2-1).
\end{equation}
Writing
\begin{equation}
    R(t)\coloneqq\int_0^t r(s)\,\mathrm ds,
\end{equation}
the solution is
\begin{eqnarray}
    \sigma&=&+1:\quad
    \zeta=-i\tan R,\quad
    \eta=-2\ln\cos R;
   \\
    \sigma&=&-1:\quad
    \zeta=-i\tanh R,\quad
    \eta=-2\ln\cosh R.
\end{eqnarray}
For $\sigma=+1$, the formula is understood on intervals where $\cos R\neq0$, with the operator relation continued by continuity.
In both cases \(\eta\) is real, so \(A=e^{-\eta/2}\) is real and
\begin{equation}
    B=ze^{-\eta/2}=-ie^{i\delta}\sin_\sigma R,
\end{equation}
where \(\sin_{+1}\equiv\sin\) and \(\sin_{-1}\equiv\sinh\). Equation~\eqref{eq:inversion} then gives
\begin{equation}
    \chi=\begin{cases}\;R&(\sigma=+1),\\\;iR&(\sigma=-1),\end{cases}
    \qquad
    \Theta_0=0,
    \qquad
    \Theta_+=e^{i\delta}R=\int_0^t\gamma(s)\,\mathrm ds,
\end{equation}
confirming the expected result in both the compact and non-compact cases. When the phase of \(\gamma(t)\) varies in time, \(\eta\) can acquire an imaginary part, \(\operatorname{Im}[e^{-\eta/2}]\neq0\), and a non-vanishing Cartan component \(\Theta_0\) is then generated generically.

\section{Krylov basis in the Schr\"odinger picture}
\label{app:sch_moving_krylov}

In this section, we show that the time-dependent Krylov dynamics is invariant under the transformation back to the Schr\"odinger picture. More precisely, the same Krylov amplitudes are obtained in a suitably chosen moving basis, provided the ladder operators are dressed by the time-dependent Cartan phases.
Let
\begin{equation}
    W(t)= U_C(t)U_E^{(I)}(t),
    \qquad
    \ket{\psi(t)}=W(t)\ket{\psi_I(t)}.
\end{equation}
We define the Schr\"odinger-picture moving Krylov basis by transporting the interaction-picture basis with \(W(t)\),
\begin{equation}
    \ket{\mathcal K_n(t)}
    \coloneqq
    W(t)\ket{K_n}.
\end{equation}
With this choice, the Schr\"odinger-picture state admits the expansion
\begin{equation}
    \ket{\psi(t)}
    =
    \sum_{n=0}^{d_K-1}\varphi_n(t)\ket{\mathcal K_n(t)},
\end{equation}
so that the expansion coefficients are exactly the same \(\varphi_n(t)\) as in the interaction picture.

The effective generator governing the amplitudes in this moving basis is
\begin{equation}
    W^\dagger(t)H(t)W(t)-iW^\dagger(t)\partial_t W(t)=H_I(t).
\end{equation}
Therefore,
\begin{equation}
    i\partial_t\varphi_n(t)
    =
    \sum_{m=0}^{d_K-1}
    \bra{K_n}H_I(t)\ket{K_m}\varphi_m(t),
\end{equation}
and, since \(H_I(t)\) is tridiagonal in the basis \(\{\ket{K_n}\}\), one recovers the same off-diagonal Krylov equation,
\begin{equation}
    i\partial_t\varphi_n(t)
    =
    \gamma(t)b_n\,\varphi_{n-1}(t)
    +
    \gamma^*(t)b_{n+1}\,\varphi_{n+1}(t).
\end{equation}
As a result, the Krylov wavefunction satisfies the same nearest-neighbor tridiagonal recursion even though the state is expressed in the Schr\"odinger picture.

It is convenient to rewrite the moving basis in terms of a moving lowest-weight state and dressed ladder operators. Defining
\begin{equation}
    \ket{\lambda_S(t)}
    \coloneqq
    W(t)\ket{\lambda}
    =
    U_C(t)U_E^{(I)}(t)\ket{\lambda},
\end{equation}
and
\begin{equation}
    \mathcal L_\pm(t)
    \coloneqq
    W(t)L_\pm W^\dagger(t),
    \qquad
    \mathcal L_0(t)\coloneqq W(t)L_0W^\dagger(t),
\end{equation}
one finds, using \([U_E^{(I)}(t),L_{\pm,0}]=0\) together with
\begin{equation}
    U_C^\dagger(t)L_+U_C(t)=e^{i\phi_\alpha(t)}L_+,
    \qquad
    U_C^\dagger(t)L_-U_C(t)=e^{-i\phi_\alpha(t)}L_-,
\end{equation}
that
\begin{equation}
    \mathcal L_+(t)=e^{-i\phi_\alpha(t)}L_+,
    \qquad
    \mathcal L_-(t)=e^{i\phi_\alpha(t)}L_-,
    \qquad
    \mathcal L_0(t)=L_0.
\end{equation}
Moreover,
\begin{equation}
    \mathcal L_-(t)\ket{\lambda_S(t)}=0,
    \qquad
    \mathcal L_0(t)\ket{\lambda_S(t)}=\lambda_\alpha\ket{\lambda_S(t)},
\end{equation}
so that \(\ket{\lambda_S(t)}\) remains a lowest-weight state for the dressed ladder structure. The moving Krylov states may therefore be written as
\begin{equation}
    \ket{\mathcal K_n(t)}
    =
    \frac{\mathcal L_+^n(t)}{\sqrt{\bra{\lambda_S(t)}\mathcal L_-^n(t)\mathcal L_+^n(t)\ket{\lambda_S(t)}}}\ket{\lambda_S(t)}
    =
    e^{-in\phi_\alpha(t)}
    \frac{L_+^n}{\sqrt{\bra{\lambda}L_-^nL_+^n\ket{\lambda}}}
    \ket{\lambda_S(t)}.
\end{equation}

Thus, the Schr\"odinger-picture basis that preserves the off-diagonal Krylov form is still generated by repeated ladder action, but now with the dressed raising operator \(\mathcal L_+(t)\), or equivalently with the bare raising operator supplemented by the compensating phase \(e^{-in\phi_\alpha(t)}\).

For comparison, suppose instead that one defines the more naive moving basis
\begin{equation}
    \ket*{\hat K_n(t)}
    \coloneqq
    \frac{L_+^n}{\sqrt{\bra{\lambda}L_-^nL_+^n\ket{\lambda}}}\ket{\lambda_S(t)}.
\end{equation}
Then
\begin{equation}
    \ket*{\hat K_n(t)}=e^{in\phi_\alpha(t)}\ket{\mathcal K_n(t)}.
\end{equation}
Expanding the Schr\"odinger-picture state as
\begin{equation}
    \ket{\psi(t)}=\sum_{n=0}^{d_K-1}\hat\varphi_n(t)\ket*{\hat K_n(t)},
\end{equation}
the amplitudes are related to the ones in the interaction picture by
\begin{equation}
    \hat\varphi_n(t)=e^{-in\phi_\alpha(t)}\varphi_n(t).
\end{equation}
Using the relation $\gamma(t)=e^{i\phi_\alpha(t)}f(t)$, inherited from the interaction-picture construction in the main text, these amplitudes satisfy
\begin{equation}
    i\partial_t\hat\varphi_n(t)
    =
    n\,\dot\phi_\alpha(t)\,\hat\varphi_n(t)
    +
    f(t)b_n\,\hat\varphi_{n-1}(t)
    +
    f^*(t)b_{n+1}\,\hat\varphi_{n+1}(t).
\end{equation}
The diagonal contribution \(n\,\dot\phi_\alpha(t)\) is thus a gauge term induced by working in the naive moving basis \(\{\ket*{\hat K_n(t)}\}\) rather than in the compensated basis \(\{\ket{\mathcal K_n(t)}\}\).

In this way, the same off-diagonal Krylov dynamics may be formulated either in the final interaction picture with the time-independent basis \(\{\ket*{K_n}\}\), or equivalently in the Schr\"odinger picture with the moving basis \(\{\ket{\mathcal K_n(t)}\}\). The two descriptions are completely equivalent and differ only in how the Cartan phase is distributed between basis states and amplitudes.

\bibliographystyle{unsrtnat}
\bibliography{references}

\end{document}